\documentclass[10pt,floats,floatfix,aps,prd,notitlepage,superscriptaddress,twocolumn,eqsecnum,nofootinbib]{revtex4-2}

\usepackage[retainorgcmds]{IEEEtrantools}

\usepackage{ifpdf}
\usepackage[utf8]{inputenc}
\usepackage[USenglish]{babel}
\usepackage{graphicx}
\usepackage{amsmath, amsthm, bm, amssymb}
\usepackage{mathrsfs}
\usepackage{multirow}
\usepackage{tikz}
\usepackage{xspace}
\usepackage{xfrac}
\usepackage{siunitx}
\usepackage[caption=false]{subfig}
\usepackage{xcolor}
\usepackage{tabularx}
\usepackage{hhline}
\usepackage{mathtools}
\usepackage{ifthen}
\usepackage{enumitem}
\usepackage{booktabs}
\usepackage[colorlinks,allcolors=blue]{hyperref}
\usetikzlibrary{positioning,calc}

\usepackage{pifont}
\newcommand{\cmark}{\ding{51}}
\newcommand{\xmark}{\ding{55}}

\def\Msun{\ensuremath{M_{\odot}}\xspace}

\def\reviewed#1{%
\ifthenelse{\equal{#1}{}}{\relax}{\noindent{\color{teal}\textbf{Reviewed: #1}}\par}}
\def\pending#1{%
\ifthenelse{\equal{#1}{}}{\relax}{\noindent{\color{red}\textbf{Pending: #1}}\par}}

\newcommand{\stacksym}[2]{\genfrac{}{}{0pt}{}{#1}{#2}}
\newcommand{\balsarapm}{\stacksym{+}{(-)}}

\def\eqref#1{Eq.~(\ref{#1})\xspace}
\def\figref#1{Fig.~\ref{#1}\xspace}
\def\tabref#1{Table~\ref{#1}\xspace}
\def\secref#1{Sec.~\ref{#1}\xspace}

\newcommand{\code}[1]{\textsf{#1}}

\newcommand{\eosgem}{\code{EOS}\xspace}
\newcommand{\eos}{EoS\xspace}

\newcommand{\cactus}{\code{Cactus}\xspace}
\newcommand{\bhah}{\code{BlackHoles@Home}\xspace}
\newcommand{\origigm}{\code{origIGM}\xspace}
\newcommand{\grhayl}{\code{GRHayL}\xspace}
\newcommand{\grhaylib}{\code{GRHayLib}\xspace}
\newcommand{\grhaylhd}{\code{GRHayLHD}\xspace}
\newcommand{\igm}{\code{IllinoisGRMHD}\xspace}
\newcommand{\grhaylhdx}{\code{GRHayLHDX}\xspace}
\newcommand{\igmx}{\code{IllinoisGRMHDX}\xspace}
\newcommand{\groovy}{\code{GRoovy}\xspace}
\newcommand{\nrpymhd}{\code{NRPyCartMHD}\xspace}

\newcommand{\etk}{\code{Einstein Toolkit}\xspace}
\newcommand{\carpet}{\code{Carpet}\xspace}
\newcommand{\carpetx}{\code{CarpetX}\xspace}
\newcommand{\grhydro}{\code{GRHydro}\xspace}
\newcommand{\clang}{\code{C}\xspace}

\newcommand{\nrpy}{\code{NRPy}\xspace}

\newcommand{\nrpyleakage}{\code{NRPyLeakage}\xspace}

\newcommand{\amrex}{\code{AMReX}\xspace}

\newcommand{\superb}{\code{superB}\xspace}

\newcommand{\etal}{\textit{et al}.\xspace}

\newcommand{\mathvar}[1]{\ensuremath{#1}}
\newcommand{\dx}{\mathvar{\Delta x}\xspace}
\newcommand{\primv}{\mathvar{\bm{P}}\xspace}
\newcommand{\consv}{\mathvar{\bm{C}}\xspace}
\newcommand{\fluxv}{\mathvar{\bm{F}}\xspace}
\newcommand{\sourcev}{\mathvar{\bm{S}}\xspace}
\newcommand{\ye}{\mathvar{Y_\mathrm{e}}\xspace}
\newcommand{\yet}{\mathvar{\tilde{Y}_\mathrm{e}}\xspace}
\newcommand{\sqrtgamma}{\mathvar{\sqrt{\gamma}}\xspace}
\newcommand{\nume}{\mathvar{n_\mathrm{e}}\xspace}
\newcommand{\numb}{\mathvar{n_\mathrm{b}}\xspace}
\newcommand{\mb}{\mathvar{m_\mathrm{b}}\xspace}

\newcommand{\Fdual}{{}^{*}\!F}

\newcommand{\orcid}[1]{\href{https://orcid.org/#1}{\includegraphics[height=\fontcharht\font`\B]{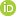}}}

\newcommand{\uidaho}{Department of Physics, University of Idaho, Moscow, ID 83843, USA}
\newcommand{\wvu}{Department of Physics and Astronomy, West Virginia University, Morgantown, WV 26506, USA}
\newcommand{\cgwc}{Center for Gravitational Waves and Cosmology, West Virginia University, Chestnut Ridge Research Building, Morgantown, WV 26505}

\begin{document}

\sloppy

\title{GRHayL: a modern, infrastructure-agnostic, extensible library for GRMHD~simulations}

\author{Samuel~Cupp~\orcid{0000-0003-1758-8376}\footnotemark}
\thanks{Equal contribution}
\email{scupp1@my.apsu.edu}
\affiliation{\uidaho}

\author{Leonardo~R.~Werneck~\orcid{0000-0002-4541-8553}}
\thanks{Equal contribution}
\email{leonardo@uidaho.edu}
\affiliation{\uidaho}

\author{Terrence~Pierre~Jacques~\orcid{0000-0002-8993-0567}}
\email{tp0052@mix.wvu.edu}
\affiliation{\wvu}
\affiliation{\cgwc}

\author{Samuel~Tootle~\orcid{0000-0001-9781-0496}}
\email{sdtootle@gmail.com}
\affiliation{\uidaho}

\author{Zachariah~B.~Etienne~\orcid{0000-0002-6838-9185}}
\email{zetienne@uidaho.edu}
\affiliation{\uidaho}
\affiliation{\wvu}
\affiliation{\cgwc}

\begin{abstract}
Interpreting multimessenger signals from neutron stars and black holes requires reliable general relativistic magnetohydrodynamics (GRMHD) simulations across rapidly evolving high-performance computing platforms, yet key algorithms are routinely rewritten within infrastructure-specific numerical relativity codes, hindering verification and reuse. We present the General Relativistic Hydrodynamics Library (\grhayl), a modular, infrastructure-agnostic GR(M)HD library providing conservative-to-primitive recovery, reconstruction, flux/source and induction operators, equations of state, and neutrino leakage through an intuitive interface. \grhayl refactors and extends the mature \igm code into reusable point- and stencilwise kernels, enabling rapid development and cross-code validation in diverse frameworks, while easing adoption of new microphysics and future accelerators. We implement the same kernels in the \etk (\carpet and \carpetx) and \bhah, demonstrating portability with minimal duplication. Validation combines continuous-integration unit tests with cross-infrastructure comparisons of analytic GRMHD Riemann problems, dynamical Tolman--Oppenheimer--Volkoff evolutions, and binary neutron star mergers, showing comparable or improved behavior over legacy \igm and established \etk codes.
\end{abstract}

\maketitle

\section{Introduction}
\label{sec:introduction}

Over the past decade, gravitational wave (GW) astronomy has experienced rapid growth in scientific output, starting with the first direct detection of GWs~\cite{GW150914} and now encompassing over 200 compact binary coalescence detections~\cite{GW151226, GW170104, GW170608, GW170814, GW170817, GW190412, GW190521, GW190521_analysis, GW190814, GW200105, LIGO_catalog1, LIGO_catalog2, LIGO_catalog3, LIGO_catalog4, LIGO_catalog5}. Among these first observations, perhaps the most impactful was GW170817/GRB170817A/AT2017gfo~\cite{GW170817}, the only multimessenger binary neutron star (BNS) merger detection to date. The GW signal was accompanied by observations of $\gamma$ rays and months-long emission throughout much of the rest of the electromagnetic (EM) spectrum~\cite{GW170817_GRB, Goldstein_2017, Tanvir_2017}. This event is expected to be just the first in GW-anchored multimessenger astronomy, which combines data from GW signals with data from coincident EM and, for sufficiently nearby sources, possibly neutrino detections to investigate some of the most extreme events in the cosmos. Looking forward, iterative improvements to the Advanced LIGO detectors, as well as future ground-based~\cite{CosmicExplorer1, CosmicExplorer2, punturo2010einstein} and space-based~\cite{LISA_2023, Luo_2016} GW detectors, will collectively provide remarkably improved detection capabilities, triggering neutrino and electromagnetic follow-ups across the spectrum.

Beyond BNS mergers, multimessenger observations of black hole--neutron star systems, circumbinary disks, core-collapse supernovae, and other compact-object environments are poised to greatly expand our understanding of high-energy astrophysical systems. However, these insights are gained by comparing observations with theoretical models, which in turn will be limited by how faithfully they track the underlying physics.

Generating self-consistent models of important multimessenger sources is exceedingly challenging, as it, in principle, requires fully solving the equations of general relativity for the spacetime, combined with a magnetohydrodynamic (MHD) treatment of turbulent magnetized plasmas, nuclear equations of state (\eos), neutrino and EM radiation transport, r-process nucleosynthesis via nuclear reaction networks, and detailed EM ray tracing. Fully self-consistent modeling of all of these processes in a single three-dimensional simulation exceeds the computational resources currently available to the community, not to mention that the relevant length and timescales span several orders of magnitude, making even one such calculation prohibitively expensive. As NR codes are extended to incorporate additional microphysics and radiation effects, the underlying frameworks have historically counterbalanced the growing computational cost by taking advantage of advances in central processing unit (CPU) architectures. However, just as the revolution in multimessenger astronomy has greatly increased the importance of improving and extending the physical realism of NR models, the rapid shift toward heterogeneous high-performance computing (HPC) systems and reliance on accelerators like graphics processing units (GPUs)---heavily driven by the advent of artificial intelligence and machine learning---now require that code frameworks be upgraded to take full advantage of state-of-the-art HPC resources.

Transforming existing CPU-tuned NR frameworks to be performant and scalable on heterogeneous CPU/GPU HPC clusters typically requires a substantial redesign of the underlying software infrastructure. As a result, significant effort has been expended in recent years to migrate existing simulation software to computational frameworks that explicitly support heterogeneous architectures and offer performance portability across devices. Many groups are working toward extending their in-house infrastructures~\cite{Ishi_2019} and/or developing entirely new ones~\cite{ET_2023_11} that build on performance-portable programming models or modern adaptive mesh refinement (AMR) infrastructures in order to fully exploit current and future HPC resources.

Two notable examples of technologies enabling this transition are \code{Kokkos}~\cite{kokkos} and \code{AMReX}~\cite{amrex}, which play complementary roles. \code{Kokkos} is a general-purpose performance-portability library that provides an abstraction layer for expressing parallelism and memory access patterns in a way that can be mapped efficiently onto diverse heterogeneous HPC architectures. In the NR community, \code{Kokkos} has been used to extend \code{Athena++}~\cite{Stone_2020, GR_Athena} to be GPU capable. The Parthenon framework~\cite{Grete:2023} further develops the original \code{Athena++} code into a more general framework for multiphysics code development on top of \code{Kokkos}. Several other projects connected to the Los Alamos National Laboratory also provide open-source components built on \code{Kokkos} that are directly relevant to NR, such as the portability abstraction library \code{Ports-of-Call}~\cite{PortsOfCall}, the tabulated data interpolation library \code{Spiner}~\cite{miller2022notquite, Miller2022}, and the \eos library \code{Singularity EOS}~\cite{miller2024singularity}.

In parallel, \code{AMReX} provides a general framework for implementing block-structured adaptive mesh refinement (AMR) codes on heterogeneous HPC systems. It supplies the mesh-management, load-balancing, and parallelization infrastructure needed to build scalable AMR applications on modern CPU/GPU clusters, and has been used by several groups to develop NR and NR-adjacent codes for these systems. The \etk~\cite{ET_2023_11}, primarily built on the \cactus framework~\cite{Cactus} and the \carpet AMR driver, is one example of an existing code base undergoing such a modernization. Recently, the new \code{AMReX}-based \carpetx code was developed and released in the \etk as a GPU-capable replacement for \carpet. Another notable example is \code{GRChombo}~\cite{Clough_2015, Andrade2021}, whose developers are in the process of creating \code{GRTeclyn} and porting the science code to the new infrastructure.

As all these efforts show, new infrastructures are required to fully leverage current and future developments in HPC and numerical techniques. Thus, in order to keep pace with these changes, there is a clear and urgent need to minimize the time and effort required to migrate NR codes between infrastructures. The current paradigm often requires significant code rewrites or refactoring, which is undesirable, as most general relativistic magnetohydrodynamics (GRMHD) algorithms are independent of the infrastructure.

Several groups have released code packages that partially address this issue. For instance, several open-source codes provide functions for conservative-to-primitive recovery schemes~\cite{Noble:2005gf, Siegel:2017jug} and interpolators for finite-temperature \eos tables~\cite{OConnor:2009iuz, Schneider:2017tfi}. While these codes require some work to incorporate into a new code, they do provide core algorithms in a reasonably self-contained form. Additionally, several libraries have been developed to provide implementations of \eos~\cite{miller2022notquite, Miller2022, miller2024singularity}, conservative-to-primitive recovery schemes~\cite{Kastaun_2021, Roland_2024}, and neutrino physics~\cite{OConnor:2014sgn, Cheong:2023fgh}.

These groups have demonstrated both the success of and demand for infrastructure- and hardware-agnostic implementations of the core GRMHD algorithms needed for NR simulations. To more comprehensively address the needs of GRMHD simulations, we introduce the General Relativistic Hydrodynamics Library (\grhayl). This library contains all the core elements required for a GRMHD simulation and represents a refactoring and extension of the \igm code~\cite{Duez_IGM, Etienne_IGM}---including the addition of tabulated \eos and neutrino leakage support in the forked code by Werneck~\etal~\cite{Werneck_IGM}---into a modular, infrastructure-agnostic library. \grhayl provides nearly all of the code required to develop an \igm-like code via point- or stencilwise library functions, allowing for quicker code development, easier cross-code and cross-infrastructure collaboration, and cleaner implementations of GRMHD codes. There is also ongoing work to add full GPU support.

This paper is organized as follows. Section~\ref{sec:library_features} discusses the individual code modules within the library and provides an overview of the available functionality. Section~\ref{sec:code_improvements} lists all the functional changes and improvements we have made to the \igm code while developing the library. Section~\ref{sec:infrastructures} describes the codes we have developed using \grhayl as well as the infrastructures themselves, in preparation for Sec.~\ref{sec:code_tests}, which compares data from the original \igm code and various new implementations using \grhayl. Finally, in Section~\ref{sec:concluding_remarks} we summarize the library features and discuss planned extensions of \grhayl's capabilities and implementations in additional infrastructures. In addition, we include two appendixes. Appendix~\ref{sec:basic_equations} briefly describes the GRMHD evolution equations and our choice of variables used for evolving them. Appendix~\ref{app:CI} details the automated, function-level continuous-integration testing that we perform to validate the library for each change made to the code repository.

\section{Library Features}
\label{sec:library_features}

The main purpose of \grhayl is to provide an infrastructure-agnostic, modular, and easily extensible GRMHD library. A key step in achieving modularity is the identification of the code's core algorithms, which in the context of GRMHD include primitive recovery routines, primitive variable reconstruction, and computation of the flux and source terms of the GRMHD equations. It is important to note that the \eos is unique in that many parts of the code depend on it directly, so special care is required to expose it through a clean, modular interface.

As part of developing \grhayl, several improvements have been made to the original \igm code. To clarify which version we are referencing, we introduce the following terminology. The original \igm and the Werneck~\etal fork are collectively referred to as \origigm, and these are the versions of \igm that have been incorporated into \grhayl. We refer to the new \grhayl-based code as \igm.

Regarding extensibility, \grhayl was designed to facilitate the seamless integration of new modules (or ``gems''), such as neutrino physics, without disturbing the existing framework. Notably, the separation is such that if one were to replace \grhayl's conservative-to-primitive routines with an alternative, the rest of the code would remain unaffected. This is a fundamental aspect of \grhayl's design, offering the flexibility for users to either extend \grhayl with their own custom routines or directly incorporate new routines into their code while leveraging the library's full capabilities.

The core part of the library---the ``chalice'' of \grhayl---is the glue that ties all the separate modules together. This must be as minimal as possible to meet the aforementioned requirements of the library. \grhayl's core contains definitions of \clang structs and simple helper functions for packing and unpacking data to and from these structs, as well as performing basic operations dependent purely on struct quantities [e.g., computing $g_{\mu\nu}$ and $g^{\mu\nu}$ from the Arnowitt-Deser-Misner (ADM) variables].

Affixed to this chalice are the various gems that implement specific features. The currently provided gems are \code{Atmosphere}, \code{Con2Prim}, \eosgem, \code{Flux\_Source}, \code{Induction}, \code{Neutrinos}, and \code{Reconstruction}. The \code{Atmosphere} gem is quite simple, as it currently only contains the constant-density atmosphere prescription used by \origigm. Other prescriptions, such as an atmosphere where hydrodynamic quantities follow a radial falloff, will be added in the future. We discuss the other gems in the rest of this section.

\subsection{\eosgem Gem}
\label{sec:eos}

The \eosgem gem provides \eos functionality within \grhayl. It currently supports hybrid piecewise polytropic \eos and includes infrastructure for fully tabulated \eos. Tabulated \eos support is presently under development and is not yet intended for production simulations. Many other gems depend on functions provided by the \eosgem gem, but this dependency has been abstracted to allow users to implement their own \eos driver without changing the other gems. The \eos information is conveyed primarily via an \eosgem struct, in addition to several function pointers that are set during initialization. Users are free to swap the \eosgem functions with their own by setting these function pointers. Further, other \eos types or implementations can be easily added to the gem. For example, providing support for the \code{Singularity EOS} library would simply require defining the proper wrapper functions that convert the expected arguments for the \grhayl functions and making the appropriate call to \code{Singularity EOS}.

A hybrid \eos~\cite{1993A&A...268..360J} assumes that the pressure and specific internal energy can be split into cold and thermal components, i.e.,
\begin{align*}
  P(\rho) &= P_{\rm cold}(\rho) + P_\mathrm{th}(\rho)\;,\\
  \epsilon(\rho) &= \epsilon_\mathrm{cold}(\rho) + \epsilon_\mathrm{th}(\rho)\;.
\end{align*}
In \grhayl, the cold pressure is represented by a piecewise polytrope (see, e.g.,~\cite{Read:2008iy}),
\begin{equation*}
  P_{\rm cold}(\rho)
  =
  \left\{
  \begin{matrix*}[l]
    K_{0}\rho^{\Gamma_{0}} &,%
    & \text{if } \rho\leq\rho_{0}\;, \\
    K_{1}\rho^{\Gamma_{1}} &,%
    & \text{if } \rho_{0}<\rho\leq\rho_{1}\;, \\
    \hspace{0.4cm}\vdots & & \hspace{1cm}\vdots\\
    K_{n}\rho^{\Gamma_{n}} &,%
    & \text{if } \rho>\rho_{n-1}\;,
  \end{matrix*}
  \right.
\end{equation*}
where $\bigl(K_{0},\{\Gamma_{i}\},\{\rho_{i}\}\bigr)$ are user-provided parameters. The cold specific internal energy is determined via the integral
\begin{equation*}
  \epsilon_\mathrm{cold}(\rho) = \int d\rho'\,\frac{P_{\rm cold}(\rho')}{\rho'^{2}}\;,
\end{equation*}
assuming that $\lim\limits_{\rho\to0}\epsilon_\mathrm{cold}=0$. The thermal component of the \eos is determined using
\begin{equation*}
P_\mathrm{th}= \bigl(\Gamma_\mathrm{th}-1\bigr)\rho\epsilon_\mathrm{th} = \bigl(\Gamma_\mathrm{th}-1\bigr)\rho\bigl(\epsilon - \epsilon_\mathrm{cold}\bigr)\;,
\end{equation*}
where $\Gamma_\mathrm{th}$ is a constant parameter that determines the conversion efficiency of kinetic to thermal energy at shocks. The \eosgem gem additionally supports a special case of the hybrid \eos called the simple \eos.%
\footnote{This is often referred to as the ideal fluid, ideal gas, or Gamma-law \eos.}
In this case, the \eos contains a single piece with $\Gamma_0 = \Gamma_\mathrm{th}$, resulting in \mbox{$P = (\Gamma_{0}-1)\rho\epsilon$}.

\subsection{\code{Con2Prim} Gem}
\label{sec:Con2Prim}

The \code{Con2Prim} gem provides several methods for recovering the primitive variables \primv from the conservative variables \consv (defined in Eqs.~\ref{eq:prims} and \ref{eq:cons}, respectively). Because of the nonlinearity of the equations, a root-finding method such as Newton--Raphson is required. In addition, this gem provides several helper functions to compute $\consv(\primv)$, to produce initial guesses for the primitives, and to enforce bounds on \consv~\cite{Faber1,Faber2} and $\primv$.

The gem includes several primitive recovery routines, including some by Noble~\etal~\cite{Noble:2005gf,Noble:2008tm}, Font~\etal~\cite{Font_2000,Etienne_2012_3}, Palenzuela~\etal~\cite{Palenzuela_2015}, and ongoing efforts to support Newman \& Hamlin~\cite{Newman_2014} (see also~\cite{Werneck_IGM} for an entropy variation of the last two). Routines are named based on the first author of their originating paper and the dimensionality of the method (e.g., \code{Noble2D}, \code{Palenzuela1D}, etc.). The gem also provides several useful diagnostics, including the number of iterations a routine took to converge to the solution and whether or not backup routines were used.

The routines are adapted from \origigm and from~\cite{Siegel:2017jug}, noting that \origigm originally adapted the Noble~\etal\ routines from \code{HARM3D}~\cite{Noble:2005gf,Noble:2008tm} and \code{HARM3D+NUC}~\cite{Murguia-Berthier:2021tnt}. We summarize all available primitive recovery routines currently implemented in \grhayl in \tabref{tab:c2p_summary}.

\begin{table}[ht]
  \centering
  \setlength{\abovecaptionskip}{10pt}
  \def\VSpace{1ex}
  \caption{List of provided conservative-to-primitive routines and their compatibility with the different \eos options.}
  \begin{tabular}{lcc}\toprule
  \multicolumn{1}{c}{\multirow{2}{*}[-2pt]{\bf Method}} & \multicolumn{2}{c}{\bf Equation of State}\\\cmidrule{2-3}
 & {\bf Simple} & {\bf Hybrid} \\\midrule
  \code{Noble2D}~\cite{Noble:2005gf} & \cmark & \cmark \\[\VSpace]
  \code{Noble1D}~\cite{Noble:2005gf} & \cmark & \cmark \\[\VSpace]
  \code{Noble1D\_entropy}~\cite{Noble:2008tm} & \cmark & \cmark \\[\VSpace]
  \code{Font1D}~\cite{Font_2000,Etienne_2012_3} & \xmark & \cmark \\[\VSpace]
  \code{Palenzuela1D}~\cite{Palenzuela_2015} & \cmark & \cmark \\[\VSpace]
  \code{Palenzuela1D\_entropy}~\cite{Werneck_IGM} & \cmark & \cmark \\\bottomrule
  \end{tabular}
  \label{tab:c2p_summary}
\end{table}

\subsection{\code{Flux\_Source} Gem}
\label{sec:flux_source}

The \code{Flux\_Source} gem provides functions to compute the flux and source terms for the right-hand sides (RHSs) of the evolved GRMHD variables as described in Appendix~\ref{sec:basic_equations}. These functions calculate the flux contributions to the RHSs using a finite-volume method, and the needed reconstructions can be performed using the \code{Reconstruction} gem. \code{Flux\_Source} uses \nrpy~\cite{Ruchlin:2018} to automatically generate optimized C code from the basic flux and source equations. The \code{Flux\_Source} gem currently only provides the Harten-Lax-van Leer (HLL)~\cite{Harten_1983} approximate Riemann solver, which has the form
\begin{equation}
    F^\mathrm{HLL} = \frac{c^{-}F_{r} + c^{+}F_{l} - c^{+}c^{-}\left(U_{r} - U_{l}\right)}{c^{+} + c^{-}}\;. \label{eq:HLL}
\end{equation}
Here $c^{-}$ is the minimum characteristic speed, $c^{+}$ is the maximum characteristic speed, $F_{r,l}$ are the hydrodynamic fluxes defined in \eqref{eq:fluxes}, and $U_{r,l}$ are the reconstructed values for the conserved variable $U$. The $r,l$ subscripts indicate fluid and spacetime quantities reconstructed at the right and left faces along some $i$-th flux direction.
To compute the characteristic speeds, we follow the prescription given by~\cite{HARM}, which approximates the general GRMHD dispersion relation as a quadratic in the wave speed. The wave speeds are then given by evaluating the quadratic equations
\begin{align*}
    c^{+} &= \max\left(\frac{-b \pm \sqrt{b^2-4ac}}{2a}\right)\;, \\
    c^{-} &= \min\left(\frac{-b \pm \sqrt{b^2-4ac}}{2a}\right)\;,
\end{align*}
with coefficients given by
\begin{align*}
a &= \left(1-v_{0}^{2}\right)\left(u^{0}\right)^{2} - v_{0}^{2}g^{00}\;, \\
b &= 2v_{0}^{2}g^{i0} - 2u^{i}u^{0}\left(1-v^{2}_{0}\right)\;, \\
c &= \left(1-v_{0}^{2}\right)\left(u^{i}\right)^{2} - v_{0}^{2}g^{ii}\;,
\end{align*}
where $v_\mathrm{A}$ is the Alfv\'en speed and $c_\mathrm{s}$ the sound speed. The variable $v_0$ is defined as
\begin{equation*}
v_{0}^{2} \equiv \frac{\omega^2}{k^2} = v_\mathrm{A}^{2} + c_\mathrm{s}^{2}\left(1-v_\mathrm{A}^{2}\right)\;,
\end{equation*}
where $\omega$ is the wave angular frequency and $k$ is the wave number. Since the characteristic speeds are computed at each cell interface from reconstructed variables, \grhayl returns final characteristic speeds
\begin{align*}
    c^{+} = \max\left(0,\ c^{+}_{l},\ c^{+}_{r}\right)\text{ and }c^{-} = -\min\left(0,\ c^{-}_{l},\ c^{-}_{r}\right)\;.
\end{align*}
We note that the characteristic speeds obtained from these equations are slightly overestimated, resulting in a modest increase in numerical diffusion and consequently in robustness.

Conforming to \grhayl's design principles, this gem computes the fluxes and source terms in a pointwise fashion. Reconstructed primitives, face-interpolated values of metric quantities, and spatial derivatives of metric quantities are taken as inputs, decoupling the gem from the method used to compute them. The functions within the gem are naturally divided into three subcategories: characteristic speeds, fluxes, and source terms. All of these use \eos functions to compute the specific enthalpy, with the characteristic speed functions also computing the sound speed. Further, the characteristic speeds and fluxes are each split into direction-dependent versions. Finally, we provide variations of the flux functions to handle the following four possible evolution options: hybrid \eos, hybrid \eos with entropy evolution, tabulated \eos, and tabulated \eos with entropy evolution.

\subsection{\code{Induction} Gem}
\label{sec:induction}

To preserve the divergence-free nature of the magnetic field, instead of directly evolving $\tilde{B}^{i}$ using the flux terms above for the induction equation, we evolve the vector potential $A_k$ using a staggered constrained-transport scheme~\cite{Balsara:1999, Etienne:2012, DelZanna_2003}. This guarantees that $\tilde{B}^{i} = \sqrt\gamma B^{i}$ remains divergence-free because, before each calculation of the evolution equations we compute
\begin{equation}
    \tilde{B}^i = \epsilon^{ijk} \partial_j A_k\;,
\end{equation}
where $\epsilon^{ijk} = \epsilon_{ijk}$ is the totally antisymmetric Levi-Civita symbol with $\epsilon^{xyz} = 1 = \epsilon_{xyz}$.

The \code{Induction} gem provides functions to facilitate the computation of the RHSs of the induction evolution equations in the Lorenz gauge. These are given by
\begin{align}
    \partial_{t}A_{i} &= \epsilon_{ijk}v^{j}\tilde{B}^{k} -\partial_i (\alpha \Phi - \beta^j A_j)\;, \label{eq:ind_A} \\
    \partial_{t}\tilde\Phi &=  -\partial_i (\alpha \sqrt\gamma A^i - \beta^i \tilde{\Phi}) - \lambda \alpha \tilde{\Phi}\;, \label{eq:ind_phi}
\end{align}
where the second terms in both equations are gauge terms. Here, $\Phi$ is the scalar potential, $\tilde{\Phi}=\sqrt\gamma\Phi$, and $\lambda$ is the damping factor of the `generalized Lorenz gauge condition.'

To illustrate how the nongauge terms in \eqref{eq:ind_A} are computed, let
\mbox{$\mathcal{E}_i \equiv -\epsilon_{ijk}v^j \tilde{B}^k$}. Focusing on, e.g., the $z$-component and dropping the gauge terms, we find
\begin{equation*}
    \partial_t A_z^{\left(i+\frac{1}{2}, j+\frac{1}{2}, k\right)}
    =
    -\mathcal{E}_z^{\left(i+\frac{1}{2}, j+\frac{1}{2}, k\right)}.
\end{equation*}

We then use a two-dimensional HLL approximate Riemann solver to determine $\mathcal{E}_i$ so that it is consistent with the staggering of the vector potential. As \eqref{eq:ind_A} suggests, this is most naturally done using the densitized magnetic field
$\tilde{B}^{i} = \sqrt\gamma B^{i}$. Continuing to focus on the $z$-component, we have
\begin{align}
  &\left(\mathcal{E}_z\right)^{\rm HLL} =
  \frac{
    c_{xy}^{++} \mathcal{E}_z^{\rm LL} +
    c_{xy}^{+-} \mathcal{E}_z^{\rm LR} +
    c_{xy}^{-+} \mathcal{E}_z^{\rm RL} +
    c_{xy}^{--} \mathcal{E}_z^{\rm RR}
  }{
    (c_{x}^{+} + c_{x}^{-})(c_{y}^{+} + c_{y}^{-})
  } \nonumber\\
  &\quad+
  \frac{c_{xx}^{+-}}{c_{x}^{+} + c_{x}^{-}}
  \bigl(\tilde{B}^{y}_{\rm R} - \tilde{B}^{y}_{\rm L}\bigr)
  -
  \frac{c_{yy}^{+-}}{c_{y}^{+} + c_{y}^{-}}
  \bigl(\tilde{B}^{x}_{\rm R} - \tilde{B}^{x}_{\rm L}\bigr),
  \label{eq:HLL_2d}
\end{align}
where \mbox{$c_{ab}^{mn} \equiv c_a^m c_b^n$}, $\mathcal{E}^{\rm LR}_z$ indicates a quadrant reconstruction that is left in $x$ and right in $y$; similarly ${\rm LL,RL,RR}$ label the other $xy$-quadrants. The subscripts $\rm L/R$ on $\tilde{B}^y$ denote left/right with respect to the $x$-direction, while $\rm L/R$ on $\tilde{B}^x$ denote left/right with respect to the $y$-direction. The formula for $\left(\mathcal{E}_x\right)^{\rm HLL}$ can be obtained from \eqref{eq:HLL_2d} via cyclic permutation of the indices, \mbox{$(z, x, y) \to (x, y, z)$}, and for $\left(\mathcal{E}_y\right)^{\rm HLL}$ via double cyclic permutation, \mbox{$(z, x, y)\to (y, z, x)$}.

Given the cyclic nature of the expression, the directional nature of the curl in our implementation is handled by simply providing the correct components in the input data. The required inputs are the velocities and magnetic fields reconstructed to the location of $A_i$, as well as the characteristic speeds $c^\pm$. Functions to compute the characteristic speeds are provided in the \code{Flux\_Source} gem, since the hydrodynamic fluxes also require $c^\pm$.

Having specified how the nongauge flux term is computed, we now turn to the interpolation needed for the gauge-related terms and the scalar potential. As with the hydrodynamic RHS, $\tilde{\Phi}^\mathrm{RHS}$ and the gauge contribution to $A_i^\mathrm{RHS}$ require staggered variable quantities. However, we can use interpolation instead of reconstruction methods because gauge variables are not as sensitive to shocks. The \code{Induction} gem provides interpolator functions to facilitate the calculation of these terms.

The primary difficulty with these quantities is that they are all sampled at different cell locations, which complicates the interpolation scheme. The hydrodynamic quantities are cell centered, and we denote these gridpoints as $(i,j,k)$, $(i+1,j,k)$, etc. The vector potential components are edge centered---e.g., $A_x$ is evaluated at the point $(i,j+1/2,k+1/2)$---and the scalar potential is vertex centered. This staggering is summarized in Table~\ref{tab:staggering}.

\begin{table}[ht]
  \centering
  \setlength{\tabcolsep}{8pt}
  \def\VSpace{1ex}
  \caption{Staggering of vector and scalar potentials relative to the cell-centered hydrodynamic quantities at $(i, j, k)$.}
  \begin{tabular}{cccc}\toprule
   {\bf Variable} & $\bm{x}$ & $\bm{y}$ & $\bm{z}$ \\\midrule
   $A_x$ & $i$ & $j+\frac{1}{2}$ & $k+\frac{1}{2}$ \\[\VSpace]
   $A_y$ & $i+\frac{1}{2}$ & $j$ & $k+\frac{1}{2}$ \\[\VSpace]
   $A_z$ & $i+\frac{1}{2}$ & $j+\frac{1}{2}$ & $k$ \\[\VSpace]
   $\tilde{\Phi}$ & $i+\frac{1}{2}$ & $j+\frac{1}{2}$ & $k+\frac{1}{2}$\\\bottomrule
  \end{tabular}
  \label{tab:staggering}
\end{table}

For $\tilde{\Phi}^\mathrm{RHS}$ and the remaining term in $A_i^\mathrm{RHS}$, we need to interpolate the vector potential to all the various staggered positions. We also need to interpolate the lapse and the shift. Finally, we need the metric to raise the index of $A_i$.

This dependency on the metric leads to several choices in how to compute the RHS. The original \origigm code uses the Baumgarte-Shaapiro-Shibata-Nakamura (BSSN) metric, but it can be more convenient in some codes to use the ADM metric. Additionally, different infrastructures assume the spacetime has different centerings. For example, the \carpet driver in the \etk assumes all variables have the same centering, so the spacetime is implicitly cell centered (i.e., it is on the same grid as the hydrodynamic variables), while \carpetx supports different grid types and has vertex-centered spacetime variables but cell-centered hydrodynamic variables. Thus, we currently provide interpolators using the ADM metric with cell- and vertex-centered variables, and the BSSN metric with cell-centered variables.

These choices only affect spacetime variable inputs, but the outputs are the same, returning all the interpolated quantities necessary to compute the remaining RHS terms. For the gauge terms in \eqref{eq:ind_A}, the interpolated quantity is all that is needed, and the user can then compute the spatial derivative to the desired order using a standard finite-difference algorithm. We also provide all the interpolated quantities to compute the RHS in \eqref{eq:ind_phi}. In addition, we provide a function for computing the RHS of \eqref{eq:ind_phi} that automatically implements upwinding in the spatial derivative of $\beta^{i}\tilde{\Phi}$, consistent with the usual BSSN shift upwinding.

\subsection{\code{Reconstruction} Gem}
\label{sec:reconstruction}

To use the HLL~\cite{Harten_1983} flux solver described in \secref{sec:flux_source}, the primitive variables must be interpolated to cell faces without introducing oscillatory behavior. The \code{Reconstruction} gem provides several methods to perform this interpolation. The library offers minmod, monotonized-central, and superbee reconstruction, which apply total variation diminishing linear interpolations~\cite{TVD} to the primitives. We refer to these methods collectively as the piecewise linear methods, noting that they are at best second-order accurate for smooth underlying solutions and revert to first order at shocks.

\grhayl also provides the piecewise parabolic method (PPM)~\cite{PPM1, PPM2}, which is third-order accurate in smooth regions and first order accurate at extrema. We note that, other than PPM, all the reconstruction functions have the same argument lists, making switching between methods very simple. The PPM scheme requires additional inputs for its special handling of shocks and thus does not use the same interface. \grhayl initializes all PPM parameters to standard values by default, but the user can tune these parameters to achieve different behavior.

\subsection{\code{Neutrinos} Gem}
\label{sec:neutrinos}

\grhayl adopts what is perhaps the most popular approach for modeling neutrino physics in GRMHD simulations: a leakage scheme~\cite{vanRiper:1981mko,Ruffert:1995fs,Rosswog:2002rt,Rosswog:2003rv,Sekiguchi:2010ep,Sekiguchi:2011zd}. In this type of scheme, experimental data are used to parametrize analytic formulas for the neutrino emission and cooling rates in terms of the optical depths and opacities, resulting in a computationally inexpensive algorithm~\cite{OConnor:2009iuz,Ott:2012kr,Neilsen:2014hha,Radice:2016dwd,Siegel:2017jug,Endrizzi:2019trv,Murguia-Berthier:2021tnt,Werneck_IGM}. However, this scheme neglects the absorption of neutrinos and fails to account for heating and lepton number changes within hot ejecta from compact-object mergers~\cite{Just:2021cls}. To account for these effects, more sophisticated techniques, like radiation transport with an M1 closure~\cite{1981MNRAS.194..439T,Shibata:2011kx} (see also~\cite{Richers:2020ntq,Radice:2021jtw}) or Monte Carlo methods~\cite{Miller:2019gig,Miller:2019dpt,Foucart:2020qjb,Foucart:2021mcb,Foucart:2021ikp}, are necessary.

The implementation of the leakage scheme in \grhayl is provided by \nrpyleakage~\cite{Werneck_IGM}.%
\footnote{We note that the neutrino source term for $\tau$ shown in Eq.\ (21) of~\cite{Werneck_IGM} is incorrect, as it is missing a factor of the lapse function. Our earlier implementation used the form \mbox{$\sqrt{\gamma} \alpha u^{0} \mathcal{Q} =  \sqrt{\gamma} W \mathcal{Q}$}, but the correct contribution is \mbox{$\alpha \sqrt{\gamma} W \mathcal{Q}$}. This typo was also present in our implementation of the source term, but has since been fixed.}
Following~\cite{Ruffert:1995fs,Galeazzi:2013mia,Siegel:2017jug,Murguia-Berthier:2021tnt,Burrows:2004vq}, we consider neutrino production from $\beta$-processes (electron/positron absorption by protons/neutrons), electron-positron pair annihilation, transverse plasmon decay, and nucleon-nucleon bremsstrahlung. The inverse reactions contribute to the total neutrino transport opacities, with inverse $\beta$-processes providing the dominant contribution to the optical depths of electron neutrinos and neutral-current scattering off neutrons the dominant contribution to the optical depths of heavy-lepton neutrinos.

For computing the optical depths, we adopt the general-purpose method of Nielsen~\etal~\cite{Neilsen:2014hha} (see also~\cite{Siegel:2017jug,Murguia-Berthier:2021tnt,Werneck_IGM}), in which neutrinos leave the system by following the path of least resistance. Refer to Werneck~\etal~\cite{Werneck_IGM} for more details on \nrpyleakage.

Given that the \code{EOS} gem's tabulated \eos support is currently experimental (and not covered in this paper), it is not recommended as the basis for production simulations. Nevertheless, the \code{Neutrinos} gem remains suitable for production use when interfaced with an external \eos implementation or infrastructure that provides the needed thermodynamic quantities.

\section{Improvements over \code{\MakeLowercase{orig}IGM}}
\label{sec:code_improvements}

Thus far, we have primarily discussed the conversion of the core \origigm routines into an infrastructure-agnostic library. In addition to simplifying and improving some of its core algorithms, we also found many ways to improve the code that ties these algorithms together when implementing \igm. As a reminder, we use \origigm to refer collectively to the original \igm thorn and the enhanced fork with neutrino leakage and experimental tabulated \eos support, while \igm refers to the new version of the thorn that uses \grhayl. In addition, we introduce the new \grhaylhd thorn, which provides a purely hydrodynamic evolution code using \grhayl.

\subsection{Code Simplification}
\label{subsec:igm_simple}

Over time, repeated inheritance of code has led to the buildup of technical debt within \origigm. While refactoring the code to write \grhayl, we found several places where the code could be simplified or condensed, with most of these improvements affecting the conservative-to-primitive routine and its surrounding functions.

One major source of unnecessary complication is the conversion of variables during primitive recovery. \origigm uses the \code{Noble2D} primitive recovery routine, which it borrowed from the \code{HARM} code. Because of this, users will find three sets of variables in \origigm: its own, \code{HARM}'s, and those required by the \code{Noble2D} routine. The code converts variables from \origigm to \code{HARM}, and then from \code{HARM} to \code{Noble2D}, performing the opposite conversions after the conservative-to-primitive routine finishes. However, many of these conversions are effectively ``do-nothing computations,'' as \origigm and \code{Noble2D} share some variables that \code{HARM} does not. We have condensed this to a single variable conversion, removing several layers of interfaces in the process.

We have also cleaned up many auxiliary functions, such as the function that enforces the inequalities that $\tilde{\tau}$ and $\tilde{S}_i$ must satisfy in order to be physical. While not individually significant, these minor improvements simplify the logical flow of the code. As discussed in Appendix~\ref{app:CI}, all of these changes are validated against the original code to ensure that their behavior is identical.

\subsection{\eos Feature Extensions}
\label{subsec:igm_improvements}

We have also made several improvements and extensions to the latest version of \igm in the \etk. The remaining subsections describe improvements to the evolution codes that implement \grhayl, and not the core library itself. This section details the extensions to \igm's \eos support. These include full support for hybrid polytropic \eos, infrastructure to incorporate tabulated \eos and neutrino leakage into the official release version of \origigm, and the addition of the simple \eos.

\origigm originally only supported simple polytropic hybrid \eos. While it had much of the code needed to extend to piecewise polytropic \eos, elements of the code contained implicit assumptions regarding the \eos. Recent improvements have added tabulated \eos~\cite{Werneck_IGM} in a forked version of \origigm. With \grhayl, we fully incorporate the piecewise polytropic hybrid \eos feature into the release version of \origigm in the \etk, and we are working on tabulated \eos support. We also add the extra features necessary to use the simple \eos. While this is of less interest for physical simulations, the commonly used Balsara tests~\cite{Balsara_2001} assume this \eos, so the original \origigm code could not run those tests with the correct \eos for comparison with the exact solution.

\subsection{More Boundary Condition Options}

As a small addition, we provide a new boundary condition option. The hydrodynamic variables use copy boundary conditions by default, while outflow boundary conditions are applied to the velocities. However, the latter condition is incompatible with 1D or 2D tests, such as the Balsara tests. As such, we introduce the option to disable the outflow boundary condition to support 1D and 2D simulations.

\subsection{Con2Prim Improvements}

The most significant changes and improvements to preexisting code are in the Con2Prim implementation. \origigm uses the \code{Noble2D} routine, with the \code{Font1D} routine as the only backup. While \code{Font1D} virtually guarantees a successful inversion, it does so by setting \mbox{$P=P_{\rm cold}$}, which introduces spurious cooling in the system. This means that any Con2Prim failures near the surface of the stars will cause discontinuities in the pressure due to the thermal component of the pressure suddenly disappearing wherever \code{Font1D} is applied, limiting the code's accuracy.

To address these issues, \igm supports up to four Con2Prim routines of the user's choosing. This design allows users to employ any of \grhayl's Con2Prim methods and control their ordering at run-time. For hybrid \eos with entropy advection evolution, a suggested combination is \code{Noble2D}, \code{Palenzuela1D}, \code{Noble1D\_entropy}, and finally \code{Palenzuela1D\_entropy}.

In \igm, the following strategy is adopted by default in order to recover the primitive variables:
\begin{enumerate}[topsep=4pt, itemsep=1pt, parsep=1pt]
    \item attempt each user-selected Con2Prim method once, in order, until one succeeds;
    \item if all selected methods fail, attempt \code{Font1D} (for hybrid \eos only); and
    \item if recovery still fails, perform an atmospheric reset.
\end{enumerate}
This multistage approach is designed to improve the likelihood that Con2Prim returns valid primitive variables. It also avoids the use of \code{Font1D} and atmospheric reset for as long as possible so as to not discard any thermal contributions. Note that \code{Font1D} is currently not optional in \igm, requiring the user to change the code in order to disable it.

\subsection{Improved Stability and Accuracy}
\label{sec:stability}

As we factored out algorithms from \origigm, we noticed a problematic pressure floor that is applied after Con2Prim routines, namely
\begin{equation*}
    P \geq 0.9 P_{\rm cold}\;.
\end{equation*}
This floor was phenomenologically motivated, as choosing a value that is slightly less than $P_{\rm cold}$ allows for pressure overshoots at the neutron star surface, resulting in less diffusion and better preservation of central density. However, a more careful analysis reveals that flooring the pressure to a value less than $P_{\rm cold}$ leads to imaginary sound speeds. While such a choice of pressure floor may produce results that appear ``better'' from a dissipation perspective, it leads to clearly unphysical behavior and risks (very rare) failures due to division by zero. As such, we enforce the pressure floor \mbox{$P\geq P_{\rm cold}$} instead.

Additionally, the computation of the source terms for the hydrodynamic variables has been improved. The original \origigm code first interpolates BSSN metric quantities to cell faces (which are needed for the flux calculations) using fourth-order interpolation. It then uses these face-centered values to compute the metric derivatives (used for the source terms) using a second-order finite difference. It is not immediately obvious what the order of accuracy of the result is, since the finite difference does depend on the larger stencil through the interpolated quantities. We modify \grhaylhd and \igm to use fourth-order finite differences of cell-centered ADM quantities for the metric derivatives, which shows noticeable improvements in conserving the central star density. Section~\ref{sec:tov_ns} discusses the effect in more detail.

\subsection{Memory Usage}
\label{subsec:memory}

While \origigm has proven to be a robust code, with other groups developing their own variants, the code does have some obvious places where it can be improved. One is the memory usage, which we have significantly reduced. We detail the changes below, but they ultimately result in a reduction in memory usage of over 40\%. This percentage only refers to memory usage by the \igm thorn, not the entire \etk, but it still represents a noticeable reduction of memory usage for simulations using \igm.

The first major change is the removal of all BSSN grid variables from the \origigm thorn. The original code used both ADM and BSSN grid variables, often recomputing them in several places. We eliminated dependencies on BSSN grid variables, greatly simplifying the \grhayl code and removing nonessential grid variables from \igm. We also improved the efficiency of the hydrodynamic RHS computations in \grhayl by generating optimized code via \nrpy, which removed the need for additional grid variables to store $T^{\mu\nu}$.

Another change concerns how we compute reconstructed variables. Previously, all variables were reconstructed in separate loops, stored, and then used for the RHS calculations, even though most were never reused. In practice, only the reconstructed velocities are needed again, to compute doubly reconstructed velocities at cell edges for $A_i^\mathrm{RHS}$. Inside the \igm thorn, we therefore combine reconstruction with the hydrodynamic-flux calculation and store only the reconstructed velocities.

All these changes result in a memory overhead reduction of over 40\% from \origigm to \igm. While entropy evolution and experimental tabulated \eos support do require a few additional grid variables, they still introduce far fewer variables than the number of variables that we have removed. The new \grhaylhd GRHD thorn has no magnetic variables, so it reduces the memory usage by ${\sim}\qty{80}{\%}$ compared to \origigm. These changes represent a significant reduction of memory usage and access.

\section{Code Infrastructures}
\label{sec:infrastructures}

\subsection{Einstein Toolkit}
\label{sec:einstein_toolkit}

Before discussing our numerical results, we first describe in more detail the infrastructures in which we have implemented GRMHD codes using \grhayl. The \etk~\cite{ET_2023_11} is a collection of codes for astrophysics simulations. Most of these codes are built on the \cactus framework~\cite{Cactus}, which provides a flexible infrastructure that allows many different code modules (called ``thorns'') to be compiled together. The \origigm code is one such thorn. While \cactus provides many features, the actual numerical grids, parallelism, and related low-level details are implemented by driver thorns. The driver currently used for most production-level simulations is the \carpet driver, which provides AMR, implements MPI parallelism, and handles memory allocation during the simulation.

Similarly, \carpetx is a newly released driver that builds on the \amrex framework~\cite{amrex}. One major change with this driver is the automation of ghost-zone synchronization and boundary condition application. This feature was added to \cactus in the \code{PreSync} update~\cite{Cupp_2018,Cupp_2019} but had not been implemented in many \carpet-based codes, including \origigm. \carpetx is designed with this new functionality in mind, and \carpetx-compatible codes are expected to provide the option for this more advanced interprocessor communication paradigm. Combined with the new looping and grid-function syntax, \origigm would need to be changed significantly to work with \carpetx. As such, an infrastructure-agnostic solution is highly desirable to reduce the work overhead of maintaining code for both drivers.

\grhayl provides this functionality within the \etk via the \grhaylib thorn, which directly compiles the \grhayl source code and sets up \grhayl parameters based on user inputs. This approach is more amenable to the \etk ecosystem than linking to an external library and improves ease of use within the \etk. Since \grhaylib only provides the core library functions, it is independent of the mesh driver.

We also introduce the \grhaylhd and \igm thorns, which reimplement the features of the \origigm thorn using \grhaylib. \grhaylhd implements a pure GRHD code, and \igm implements the GRMHD code. These thorns define grid variables, schedule functions with loops over the grid, and fill the data structures needed by the \grhayl functions. Everything else is external to these thorns. Thus, code duplication is minimized, and the pure hydrodynamic code benefits from the reduced computational cost and memory usage expected from removing the magnetic components.

Similarly, we provide a \carpetx-based thorn named \grhaylhdx, which again depends on \grhaylib. The \igmx thorn is also in development, with expected inclusion in a future release of the \etk. With the inclusion of \carpetx in the \etk, providing reliable GRMHD codes for the new driver is critical for future science using the \etk, with \code{AsterX}~\cite{Kalinani:2024rbk} a noteworthy example. One important caveat is that \carpetx also supports GPUs. While work to support GPUs has begun, \grhayl currently only supports CPU code. As such, the \carpetx thorns run on the host at the moment. Once the library functions are available on the device, the thorns' code can be easily switched to GPU-based computation.

\subsection{BlackHoles@Home}
\label{sec:nrpy}

\bhah is an infrastructure built on the code-generation package \nrpy~\cite{Ruchlin:2018,Etienne_nrpytutorial}. \nrpy allows for efficient generation of optimized \code{C} code from symbolic \code{Python} expressions. \bhah leverages this \code{Python} package to produce minimal but efficient numerical relativity code using high-order finite-differencing and time-stepping algorithms with extrapolation and high-order radiation outer boundary conditions. The infrastructure takes advantage of \nrpy's \code{C} code generation to produce state-of-the-art code for binary black hole simulations. The core goal of the infrastructure is to reduce the memory usage and inherent inefficiencies of box-in-box AMR on Cartesian grids by using multicoordinate, multipatch grids.

Previously, this infrastructure only supported vacuum spacetime evolutions. We have developed two \grhayl-based GR(M)HD codes for \bhah: the GRHD code \groovy and its GRMHD extension \nrpymhd. \groovy~\cite{Jacques:2024pxh} implements \grhayl within \bhah, yielding a GRHD code for fully dynamical spacetimes that supports single-patch curvilinear coordinates via a reference-metric formulation~\cite{Montero:2012yr,Baumgarte:2012xy,Mewes:2020vic}. \nrpymhd is a specialization of \groovy that extends it to GRMHD while restricting the coordinate system to Cartesian coordinates. We are currently developing a supercomputer-ready version of \groovy that leverages \code{Charm++} task-based parallelism through \superb~\cite{Jadoo:2025lgm}.

We note that the \etk thorns \code{Baikal} and \code{BaikalVacuum} for spacetime evolution are also \nrpy-generated, and \bhah uses the exact same core routines internally. This allows for more meaningful comparisons between the two infrastructures, since the spacetime evolution uses the same code.

\section{Code Tests}
\label{sec:code_tests}

We approach testing and validation through two complementary avenues. First, the core \grhayl library has many function-level tests that are regularly performed. The details of our continuous-integration testing methods are given in Appendix~\ref{app:CI}. Second, we compare physics simulations produced in several different infrastructures with results from \origigm. Through this testing, we validate the behavior of the replacements for \origigm by showing that the differences between the \grhayl-based codes and \origigm are within expectations. Additionally, we demonstrate successful parallel development of GRMHD codes using \grhayl in \bhah, \carpet, and \carpetx.

\subsection{1D Shock Tests}
\label{sec:1d_shock_tests}

Balsara et al.~\cite{Balsara_2001} provide a very challenging battery of tests using one-dimensional GRMHD flows characterized by sharp features---along with extreme Lorentz factors or magnetization---that push GRMHD codes to their limits. These tests also have analytic solutions, allowing for direct validation of a numerical code's results. While \origigm is capable of passing these tests, one must remove its pressure floors and ceilings and adjust some \eos functions in order to do so, preventing them from being run without manual changes to the code. Because of our changes and improvements, \igm can run these tests without modification by using the simple \eos option. Here we briefly describe the Balsara tests and present results comparing \igm and \grhaylhd with analytical results and \code{GRHydro}, another trusted GRMHD code in the \etk.

\begin{table*}[htb]
\centering
\setlength{\abovecaptionskip}{10pt}
\def\arraystretch{1.7}
\setlength{\tabcolsep}{5pt}
\caption{Basic setup for the left (right) side of the shock for the Balsara tests.}
 \begin{tabular}{||c|c c c c|c c c|c c c|c c||}
 \hline
 Test & $t_\mathrm{final}$ & $\Gamma$ & $\rho$ & $P$ & $\tilde{u}^x$ & $\tilde{u}^y$ & $\tilde{u}^z$ & $B^x$ & $B^y$ & $B^z$ & $W$ & $P/P_\mathrm{mag}$ \\\hhline{#=|====|===|===|==#}
 1 & $\frac{2}{5}$ & $2$ & $1$ ($\frac{1}{8}$) & $1$ ($\frac{1}{10}$) & $0$ & $0$ & $0$ & $\frac{1}{2}$ & $\balsarapm 1$ & $0$ & $1$ & $1.6$ ($0.16$) \\
 2 & \vdots & $\frac{5}{3}$ & $1$ & $30$ ($1$) & \vdots & \vdots & \vdots & $5$ & $6$ ($\frac{7}{10}$) & $6$ ($\frac{7}{10}$) & \vdots & $0.62$ ($0.077$) \\
 3 & \vdots & \vdots & \vdots & $1000$ ($\frac{1}{10}$) & \vdots & \vdots & \vdots & $10$ & $7$ ($\frac{7}{10}$) & $7$ ($\frac{7}{10}$) & \vdots & $10.1$ ($0.002$) \\
 4 & \vdots & \vdots & \vdots & $\frac{1}{10}$ & $\balsarapm 0.999$ & \vdots & \vdots & \vdots & $\balsarapm 7$ & $\balsarapm 7$ & $22.4$ & $2.01$ ($2.02$) \\
 5 & $\frac{11}{20}$ & \vdots & $\frac{27}{25}$ ($1$) & $\frac{19}{20}$ ($1$) & $\frac{2}{5}$ ($-\frac{9}{20}$) & $\frac{3}{10}$ ($-\frac{1}{5}$) & $\frac{1}{5}$ & $2$ & $\frac{3}{10}$ ($-\frac{7}{10}$) & $\frac{3}{10}$ ($\frac{1}{2}$) & $1.19$ ($1.18$) & $0.27$ ($0.34$) \\\hline
 \end{tabular}
\label{tab:balsara}
\end{table*}

The Balsara tests are all pure MHD tests in static Minkowski spacetime and have an initial shock placed at $x=0$. Note that the initial data for these tests prescribe the Valencia velocity $\tilde{u}^{i}$ instead of $v^{i}$, but in flat space \mbox{$\tilde{u}^{i}=v^{i}$} (see Eq.~\ref{eq:valencia_v}). In Table~\ref{tab:balsara} we display the initial parameters for the different Balsara tests, also listing the Lorentz factor and the ratio $P/P_{\rm mag}$.

As shown in the table, all of these tests have very small ratios of gas pressure to magnetic pressure. In addition, Balsara~3 has a pressure shock of 4 orders of magnitude without any change to the density, and the Balsara~4 test has an enormous Lorentz factor of over 22. Simply put, these tests all push an MHD code to the limits of the physics they are designed to model and serve as excellent first tests that an MHD code is working properly.

\begin{figure}[ht]
  \centering
  \includegraphics[width=\linewidth]{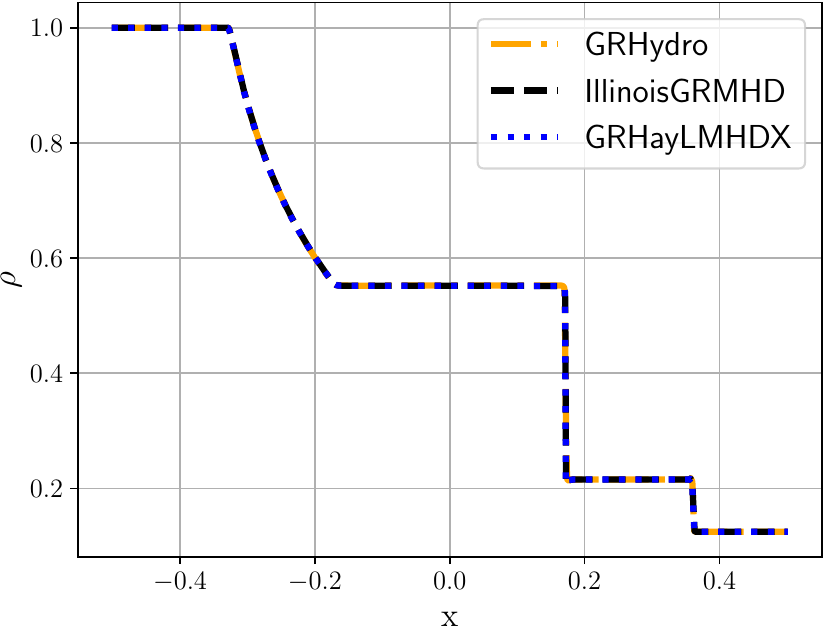}
  \caption{Comparison of \code{GRHydro}, \igm, and \grhaylhdx with the Balsara~0 test.}
  \label{fig:BalsaraX}
\end{figure}

\begin{figure}[ht]
  \centering
  \includegraphics[width=\linewidth]{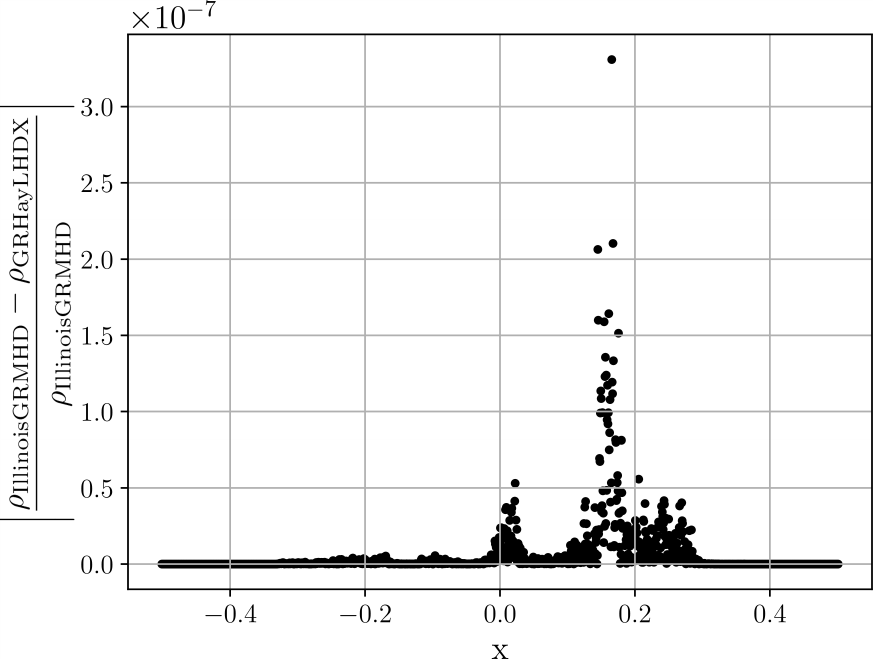}
  \caption{Relative difference between \igm and \grhaylhdx in the Balsara~0 test.}
  \label{fig:BalsaraX_diff}
\end{figure}

\begin{figure*}[ht]
  \centering
  \includegraphics[width=\linewidth]{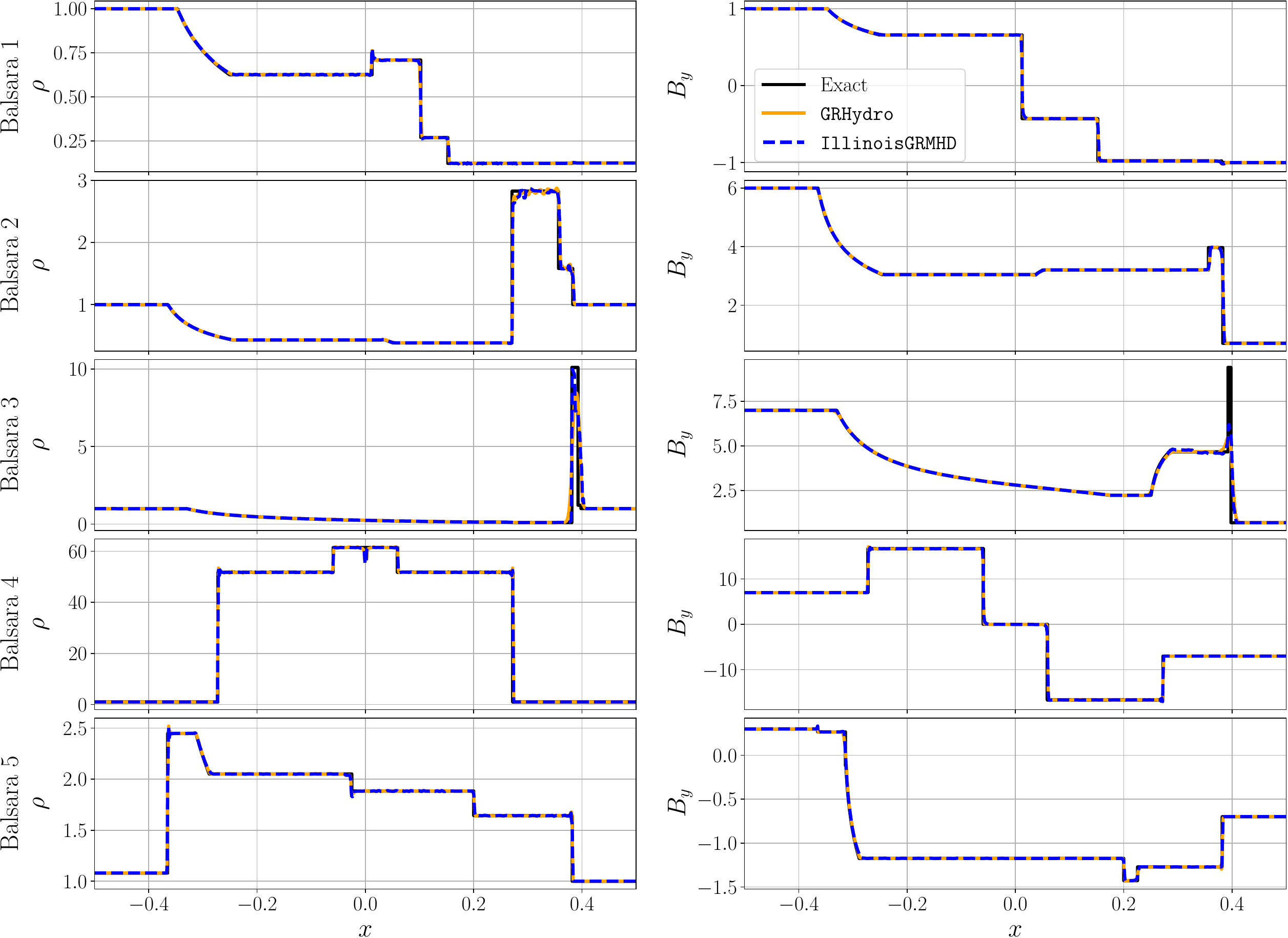}
  \caption{Comparison of \code{GRHydro} and \igm against the exact solution for the Balsara tests. The left column shows the density at the final time of the test, and the right column shows the $y$ component of the magnetic field.}
  \label{fig:Balsara}
\end{figure*}

One drawback to the Balsara tests is that they all have nonzero magnetic fields, which means that the analytical results cannot be used for purely hydrodynamic codes. However, we can still use a modified version of the Balsara tests to compare different codes. Since the shocks are still significant even in just the hydrodynamic variables, this still serves as an excellent test of the purely hydrodynamic \grhayl codes. We therefore compare and validate the hydrodynamic codes using a ``Balsara~0'' test, obtained from Balsara~1 by setting the magnetic fields to zero.

In \figref{fig:BalsaraX}, we compare results from the Balsara~0 test with \grhydro, \igm, and \grhaylhdx. All three codes show excellent agreement, validating the \carpetx-based code. Since \igm and \grhaylhdx are both based on \grhayl, the only differences are that one explicitly evolves the magnetic variables and that they are implemented in different infrastructures. As such, this provides an excellent comparison between them.

In \figref{fig:BalsaraX_diff}, we show the relative difference between the final results of the two \grhayl-based codes. The error is concentrated around the shocks or areas most affected by the shock propagation, as can be seen by examining the shock profile in \figref{fig:BalsaraX} and comparing it with the errors in \figref{fig:BalsaraX_diff}. The peak around zero represents lingering effects of the original shock, which started in the center.

As part of our validation, we also ran all five Balsara tests with \igm and \grhydro, a well-known and tested code within the \etk. All the simulations have 1600 points in the shock direction and 8 points in the perpendicular directions. Except for the Balsara~3 test, the simulations all use PPM reconstruction and the HLL approximate Riemann solver for the hydrodynamic quantities. In the Balsara~3 test, \grhydro crashes due to Con2Prim failures when using PPM, so \grhydro instead uses minmod for the reconstruction. This failure mode is not unique to \grhydro, as \code{Spritz} also uses minmod for this test. We also see Con2Prim failures near the outer edges of the grid, but our backup routines prevent the run from ending prematurely. While \grhayl provides minmod, \igm currently only supports PPM, so we cannot match \grhydro's settings for this test.

The primary difference between the two codes is in the magnetic sector. \grhydro only supports evolving the magnetic field directly or evolving a cell-centered vector potential, while \igm evolves a staggered (i.e., edge-centered) vector potential. We choose to compare against \grhydro with magnetic field evolution, which is the default and more heavily tested evolution scheme in the code. Further, it has been shown that evolution of the cell-centered vector potential leads to increased noise around shocks (see, e.g.,~\cite{Spritz}).

In addition to these numerical codes, we use the analytical solution produced by the code presented in~\cite{Giacomazzo_2006}. Figure~\ref{fig:Balsara} shows the results for the Balsara tests with \igm and \grhydro, overlaid with the exact solution. We see that the two codes agree very well not only with the exact solution, but also with each other. In the Balsara~3 test, \igm captures the shock better due to using PPM instead of minmod.

\subsection{Tolman-Oppenheimer-Volkoff (TOV) Neutron Star}
\label{sec:tov_ns}

To further validate \grhayl and its implementations within the \etk and the \bhah infrastructures, we performed simulations with single neutron stars using TOV initial data. The TOV solution is a spherically symmetric solution to Einstein's equations with a nonzero stress-energy tensor. In the following subsections we show numerical results from evolving TOV initial data in a fully dynamical, three-dimensional spacetime with no symmetry assumptions. All initial data used for these tests were generated using the TOV solver within \nrpy.

\subsubsection{Comparison across infrastructures}

Our first suite of tests consists of evolving the same TOV initial data across different infrastructures, comparing the following codes: \origigm and \igm within the \etk, and \nrpymhd. The initial central density of the star is \mbox{$\rho_{c} = 1.28\times10^{-3}$} in code units, and we adopt an initially cold hybrid \eos with \mbox{$K_{0}=100$} and \mbox{$\Gamma_{0}=\Gamma_{\rm th}=2$}. The TOV initial data do not include magnetic fields, making this a pure hydrodynamics test. The spacetime is evolved using \code{Baikal}~\cite{Ruchlin:2018, Etienne_nrpytutorial}, available within both the \etk and \bhah.

\code{Baikal} evolves the spacetime variables using the BSSN formulation~\cite{Brown_2009}, and employs the $1+\log$ lapse~\cite{Campanelli_2006} and a second-order, noncovariant, advective shift evolution~\cite{Meter_2006} as gauge conditions, i.e.
\begin{align*}
\partial_t \alpha &= \beta^i \partial_i \alpha - 2 \alpha K, \\
\partial_t \beta^i &= \beta^j \partial_j \beta^i + B^{i}, \\
\partial_t B^i &= \beta^j \partial_j B^i + \frac{3}{4} \partial_{0} \bar{\Lambda}^{i} - \eta B^{i},
\end{align*}
where $K$ is the trace of the extrinsic curvature $K_{ij}$, $\bar{\Lambda}^{i}$ and $B^i$ are auxiliary variables, and $\eta$ is the shift damping parameter. We set $\eta=1/3$, use Kreiss-Oliger dissipation at a strength of $0.2$~\cite{Kreiss_1973}, and use fourth-order finite differencing for the spacetime fields. For the hydrodynamic fields we use PPM reconstruction, the \code{Noble2D} conservative-to-primitive solver, and set the atmospheric density to $\rho_{\mathrm{atm}} = \rho_c \times 10^{-9}$. In all tests we ensure that the diameter of the star is sampled with at least ${\approx}30$ points along a coordinate axis. These codes evolve the initial data forward in time using the method of lines and a fourth-order Runge-Kutta scheme and, apart from \nrpymhd, use five levels of fixed mesh refinement (FMR) in Cartesian coordinates with the outer boundary placed at ${\pm}240$ in all three directions. We evolve the initial data at two resolutions, MR and HR, corresponding to resolutions on the coarsest level of $dx=8.0$ and $dx=5.0$, respectively. Since \nrpymhd supports only a single Cartesian grid, we place its outer boundary at ${\pm}40$ in all three directions and choose the grid spacing to match that of the finest refinement level of the FMR hierarchy.

\begin{figure}[ht]
  \centering
  \includegraphics[width=\linewidth]{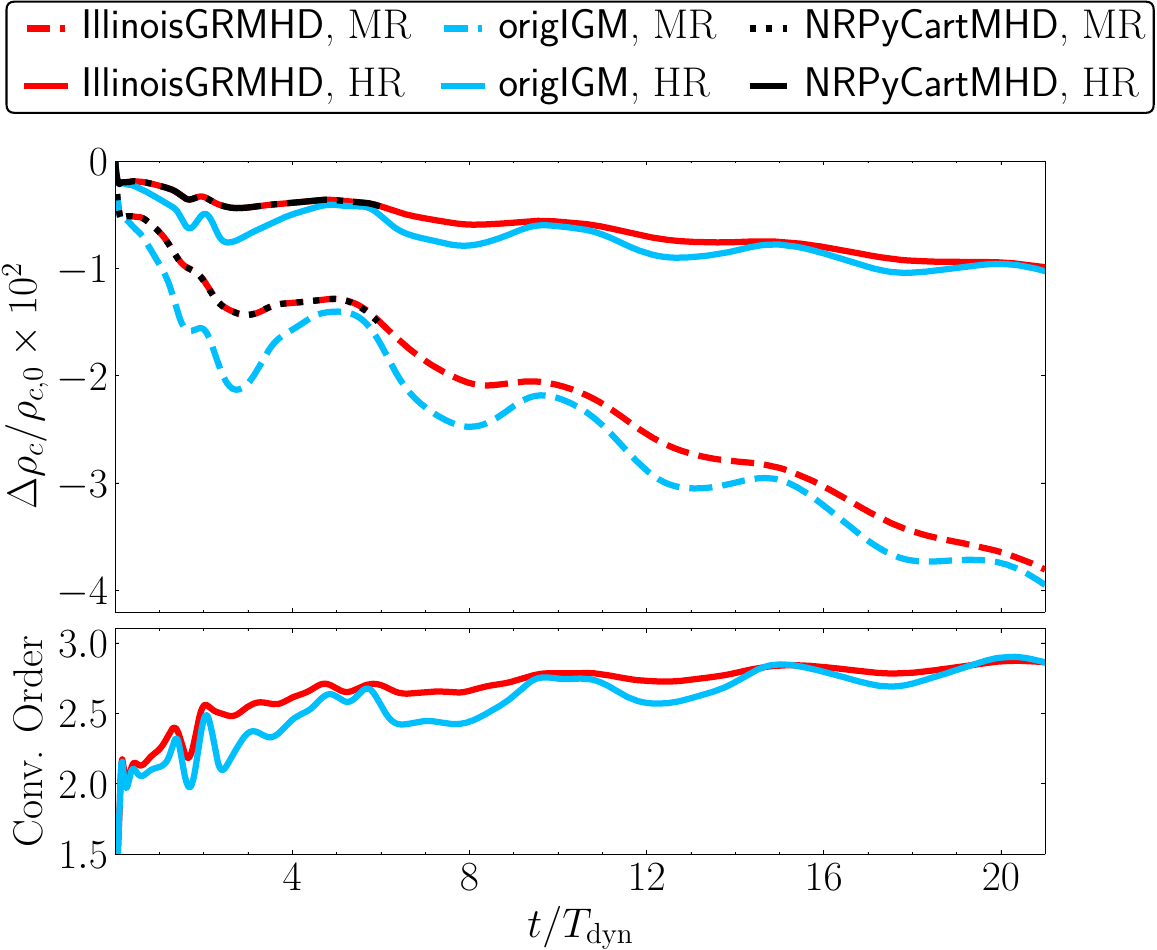}
  \caption{\textbf{Top}: comparison of nonmagnetized TOV initial data evolution between \origigm, \igm, \grhaylhdx, and \nrpymhd, each at two resolutions. \igm and \nrpymhd are equivalent until outer boundary effects are apparent, therefore we truncate the results of \nrpymhd before they manifest. \textbf{Bottom}: comparison of convergence order between \origigm and \igm. \igm consistently outperforms \origigm.}
  \label{fig:tov_compare}
\end{figure}

In the top panel of \figref{fig:tov_compare} we show results of the time evolution of the normalized change in central density for all codes, truncating the results from \nrpymhd when outer boundary effects become apparent, with all the \grhayl-based codes using the full suite of changes from \secref{sec:code_improvements}. These results demonstrate that the frequencies of the fundamental modes of oscillation for the star are recovered~\cite{Font_2000_nonlinear_tests, Font_2002}. Further, the amplitude of these oscillations, which are driven by discretization errors in the nominally static initial data, is expected to converge to zero with increasing resolution. In the bottom panel of \figref{fig:tov_compare} we plot the approximate convergence order of the error in central density for \origigm and \igm. We observe that they have a similar convergence order at around $n=2.5$, consistent with what was found in previous studies (e.g.,~\cite{Werneck_IGM}).

\begin{figure}[ht]
  \centering
  \includegraphics[width=\linewidth]{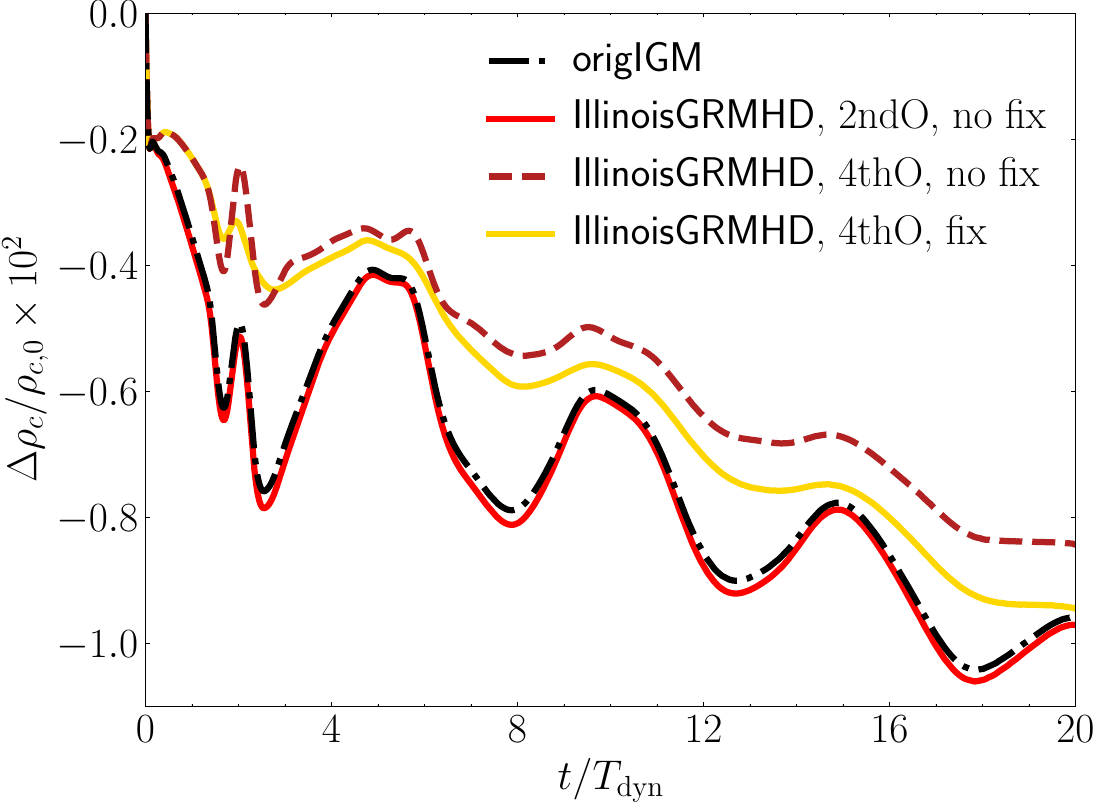}
  \caption{Comparison of nonmagnetized TOV initial data evolution between \origigm and various changes made to \igm during development.}
  \label{fig:tov_changes}
\end{figure}

In \figref{fig:tov_changes} we plot the normalized change in central density over time, comparing \origigm and different versions of \igm. These versions represent the changes described in \secref{sec:code_improvements}. The second-order data best match the \origigm code, and the central density has similar behavior and loss due to diffusion. As discussed in \secref{subsec:igm_improvements}, the local grid data needed to perform fourth-order finite differencing are already available, so we can immediately change the derivative calculation without requiring any extra data. This change better preserves the central density.

On the other hand, allowing the pressure to fall below the cold pressure leads to cooling and limits diffusion. Therefore, the change to the pressure floor results in more diffusion. As discussed in \secref{sec:stability}, increased diffusion is a necessary sacrifice to avoid unphysical behavior and possible divisions by zero.

\begin{figure}[ht!]
  \centering
  \includegraphics[width=\linewidth]{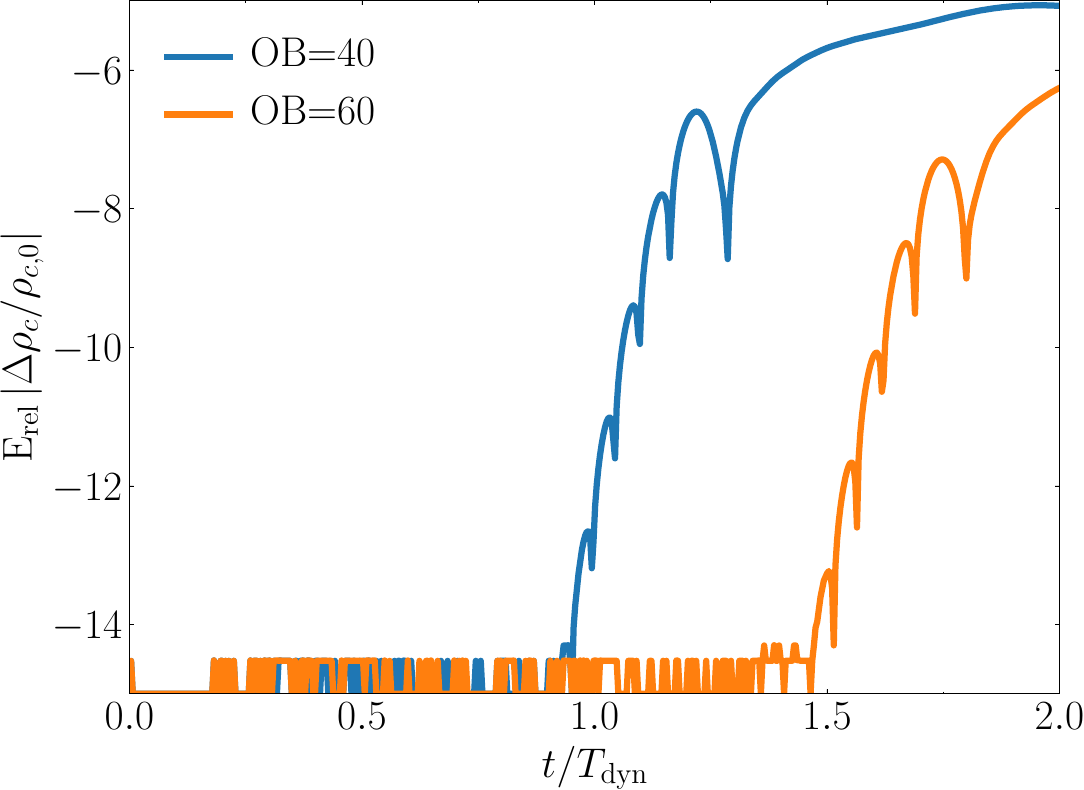}
  \caption{Relative difference between \igm and \nrpymhd, comparing two outer boundary (OB) locations while keeping the resolution fixed and using a single refinement level.}
  \label{fig:round-off_BHAH_ETK}
\end{figure}

Finally, to confirm the effective integration of \grhayl within both the \etk and \bhah infrastructures, we plot the relative difference between the two in \figref{fig:round-off_BHAH_ETK}, using a single Cartesian mesh for both. We observe that as we vary the location of the outer boundary, while keeping the resolution fixed, the duration over which we maintain round-off-level agreement is extended. This demonstrates that the observed discrepancies arise solely from the slightly different implementations of Sommerfeld radiation boundary conditions in \carpet and \bhah.

\subsection{Piecewise polytrope \eos study}

\begin{figure}[ht]
  \centering
  \includegraphics[width=\linewidth]{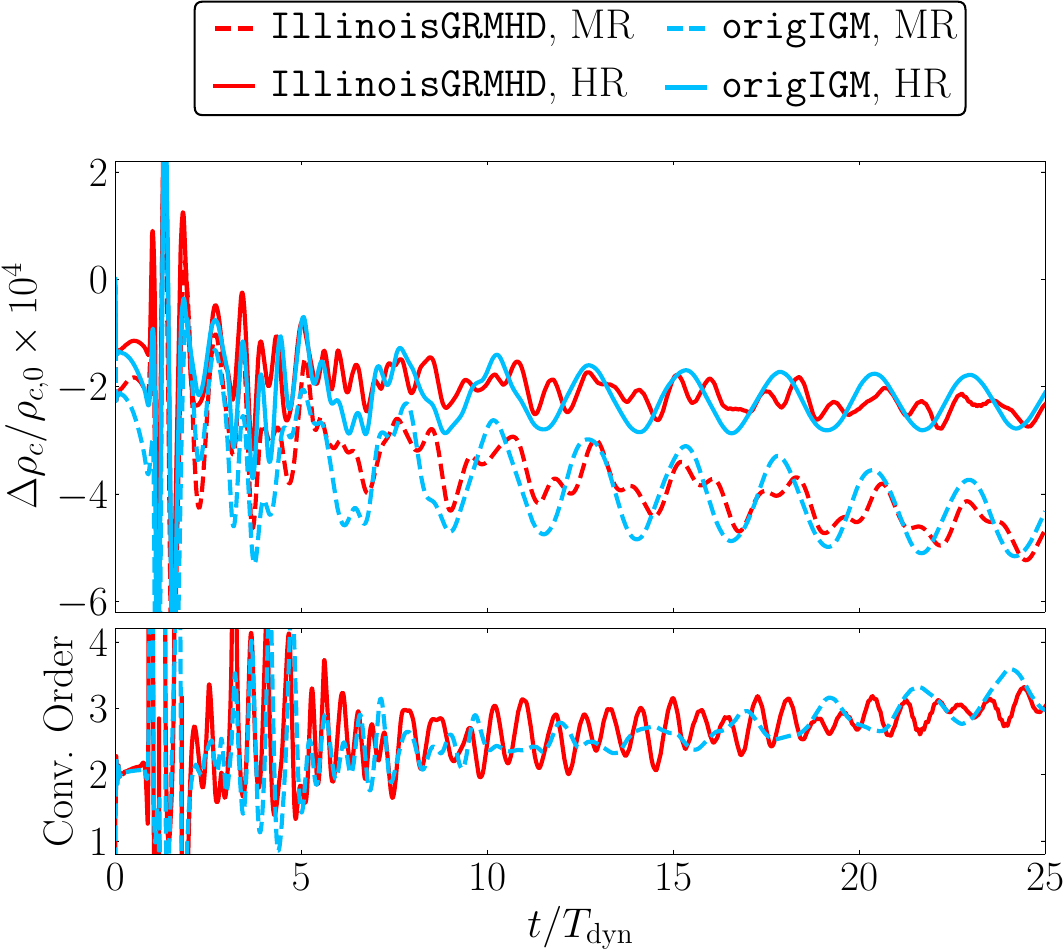}
  \caption{Same as \figref{fig:tov_compare} but for piecewise polytropic SLy \eos.}
  \label{fig:TOV_compare_SLy}
\end{figure}

\begin{figure}[ht]
  \centering
  \includegraphics[width=\linewidth]{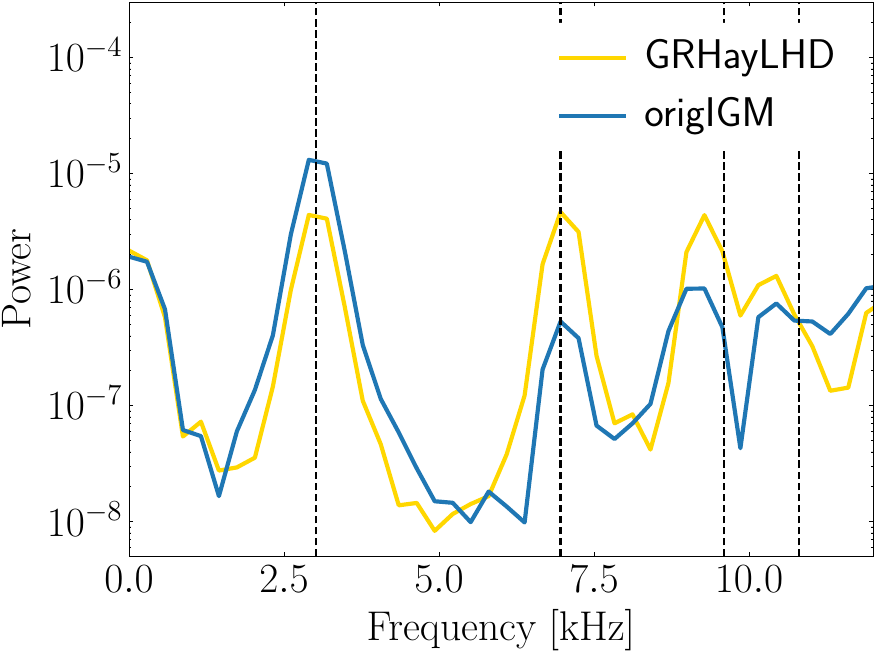}
  \caption{Fourier transform of time series of the central density evolution for a TOV in dynamical spacetime with the piecewise polytropic representation of the SLy \eos. Vertical dashed lines indicate the frequencies of the lowest four radial oscillation modes for the SLy \eos~\cite{Sen:2022kva}.}
  \label{fig:SLy_PPEOS_FFT}
\end{figure}

We next evolve TOV initial data for a $1.4 M_{\odot}$ star with the SLy \eos~\cite{Douchin:2001sv} in the \etk, adopting a piecewise polytropic~\cite{Read:2008iy} (PP\eos) representation. We sample the diameter of the star with ${\approx}130$ points, use six levels of FMR, and place the outer boundary at ${\pm}384$. Using the same spacetime settings as the previous test, we evolve the star for ${\approx}25$ dynamical timescales on medium and high-resolution grids, which have grid spacings of \mbox{$\Delta x=4.0$} and \mbox{$\Delta x=3.2$} at the coarsest refinement level, respectively. The normalized drift of the central density as a function of time for these two resolutions is shown in \figref{fig:TOV_compare_SLy}. A discrete Fourier transform (\figref{fig:SLy_PPEOS_FFT}) recovers the expected $F$- and $p_1$-mode frequencies from linear perturbation theory to better than \qty{1.5}{\percent}.

\subsection{BNS Inspiral, Merger, and Collapse of Hypermassive Neutron Star}
\label{sec:bns_inspiral}

As a final test, we consider BNS mergers, which pose a particularly challenging problem for any GRMHD code. At present, however, \groovy is implemented only in the single-curvilinear-patch version of \bhah. Although multipatch support is under development, it is not practical to perform BNS simulations within a single grid patch. The current development status of \carpetx similarly restricts the simulations we can perform: several key capabilities, including BNS initial data and various GRHD diagnostics, are not yet available in the new driver. Given these limitations and the high computational cost of BNS simulations, in this work we restrict our attention to comparisons between \origigm and \igm using \carpet.

\subsubsection{Hybrid \eos, Simple Polytrope}

\begin{figure}[ht]
  \centering
  \includegraphics[width=\linewidth]{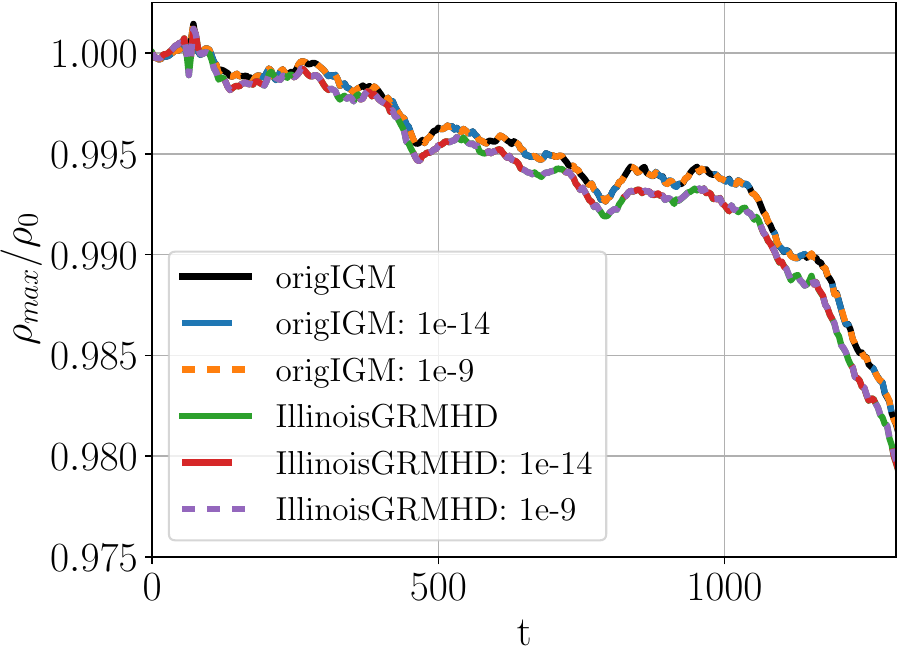}
  \caption{Comparison of the maximum density for \origigm and \igm during inspiral with both perturbed and unperturbed simulations. The $y$-axis is rescaled by the maximum density at $t=0$.}
\label{fig:inspiral}
\end{figure}

\begin{figure}[ht]
  \centering
  \includegraphics[width=\linewidth]{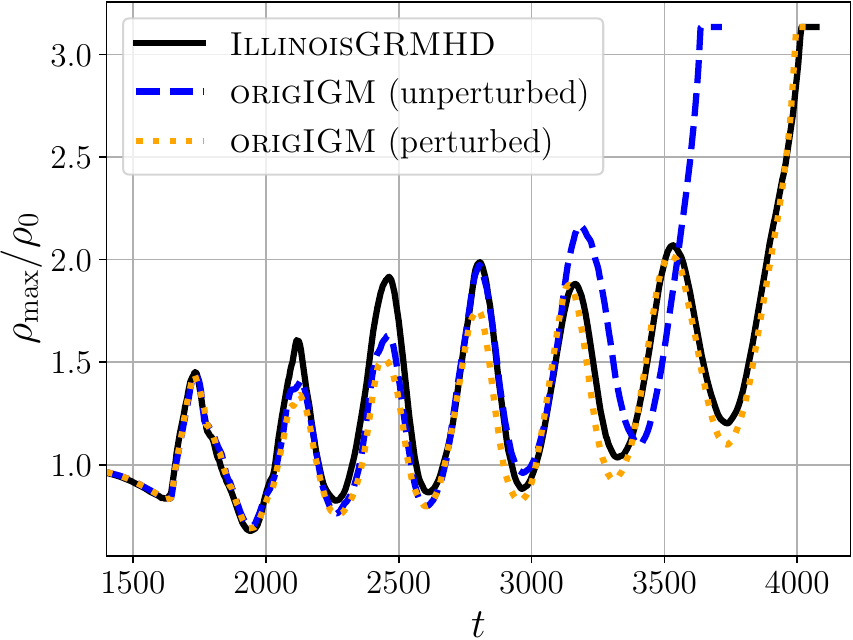}
  \caption{Comparison of the maximum density for \origigm with randomly perturbed initial data from hypermassive neutron star formation to collapse. We note that the near overlap between \igm and the perturbed run with \origigm is purely coincidental.}
  \label{fig:rmax_hns}
\end{figure}

For the hybrid \eos comparison, we use $\Gamma = \Gamma_\mathrm{th} = 2$ and $K\approx123.64$. The limits on the density are set to $\rho_\mathrm{min} = \rho_\mathrm{atm} = 10^{-13}$ and $\rho_\mathrm{max} = 0.003 = 3.1 \rho_{c,0}$, where $\rho_{c,0}$ is the initial central density of the neutron stars. The primary Con2Prim method is set to \code{Noble2D}, with backups of \code{Palenzuela1D}, \code{Noble1D\_entropy}, and finally \code{Font1D}. This does differ from \origigm, which uses the only available backup of \code{Font1D}.

We use \code{LORENE} initial data for an equal-mass BNS system with an initial separation of $45$~km. Each star has mass $M\approx1.625\,M_\odot$ and radius $R\approx\qty{13.67}{\km}$. We use the \code{Seed\_Magnetic\_Fields} thorn to set up the initial magnetic fields and the \code{Baikal} thorn to evolve the spacetime. Further details regarding the AMR grids and thorn setup are provided in the hybrid BNS parameter file packaged with the \igm thorn.

Because of the various improvements (detailed in \secref{sec:code_improvements}), the codes no longer agree at the round-off level. To assess their level of agreement, we perform three BNS simulations with both evolution codes to analyze their behavior. One is the control with no perturbations, and the other two perturb the initial \primv data at the 9th and 14th significant digits, respectively.

Figure~\ref{fig:inspiral} shows the quantity $\rho_\mathrm{max}/\rho_0$ during the inspiral phase of the simulation for all six simulations. The two codes show the same behavior, and the differences follow the same trends as the TOV simulations. At this stage of the simulation, the perturbations show no significant effect on the central density of the stars. The central densities dissipate away faster in \igm due to the change to the $P_\mathrm{min}$ floor, which has a slightly stronger effect in these simulations than in the TOV simulations.

The perturbative effects start significantly affecting the evolution during the hypermassive neutron star phase, where even small changes can affect the evolution of the system. Figure~\ref{fig:rmax_hns} compares $\rho_\mathrm{max}$ between \origigm simulations (unperturbed and perturbed at the 14th digit) and an unperturbed \igm simulation. We cut off the data once the simulation reaches the maximum allowed density for the simulation. As shown, perturbations to initial data cause the number of oscillations to change in the original \origigm code. The new \igm has round-off-level differences at every Con2Prim evaluation relative to the original code due to code improvements and changes, which would similarly cause differences between the old and new results. Despite these cumulative perturbations, \igm's results are still on the order of a round-off-level perturbation relative to the original code. Given the unstable nature of the hypermassive neutron star and the repeated perturbations caused by various code improvements and simplifications, \igm displays remarkable agreement with \origigm.

\begin{figure*}[p]
  \centering
  \setlength{\abovecaptionskip}{10pt}
  \begin{tabular}{c|c}
    \origigm & \igm \\
    \subfloat[$t=0$.]{
      \includegraphics[width=0.45\textwidth,clip]{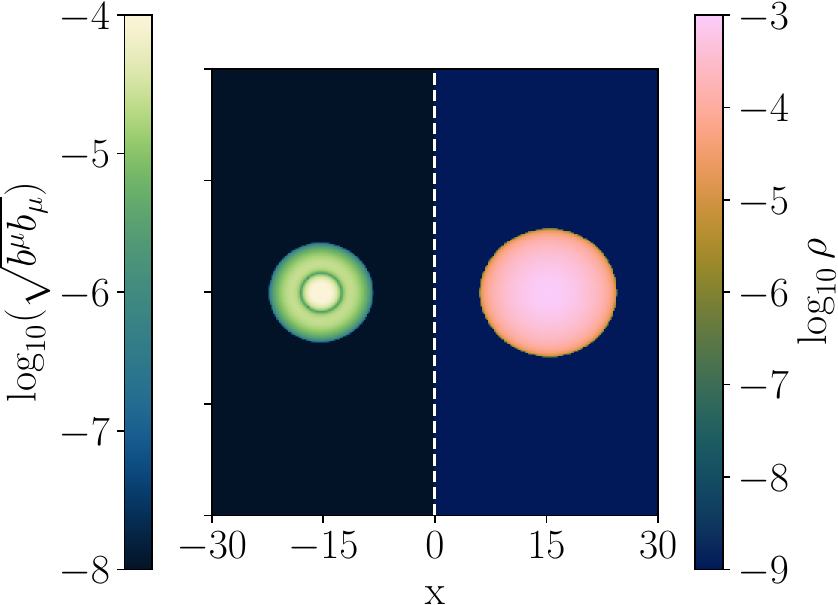}
      \label{fig:hybrid_bns_inspiral_igm1}
    }
    &
    \subfloat[$t=0$.]{
      % \addtocounter{subfigure}{2}
      \includegraphics[width=0.45\textwidth,clip]{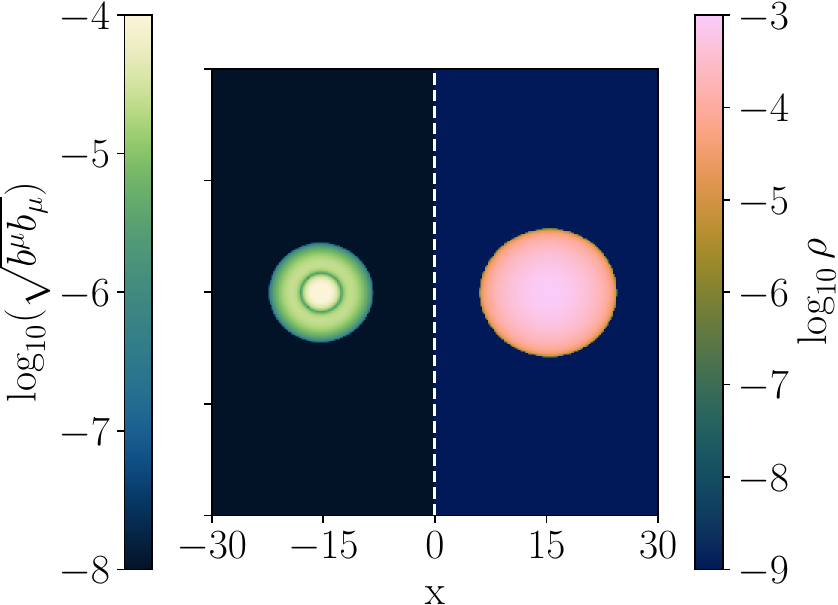}
      \label{fig:hybrid_bns_inspiral_grhayl1}
    }
    \\
    \subfloat[$t=944$ (two full orbits).]{
      % \addtocounter{subfigure}{-3}
      \includegraphics[width=0.45\textwidth,clip]{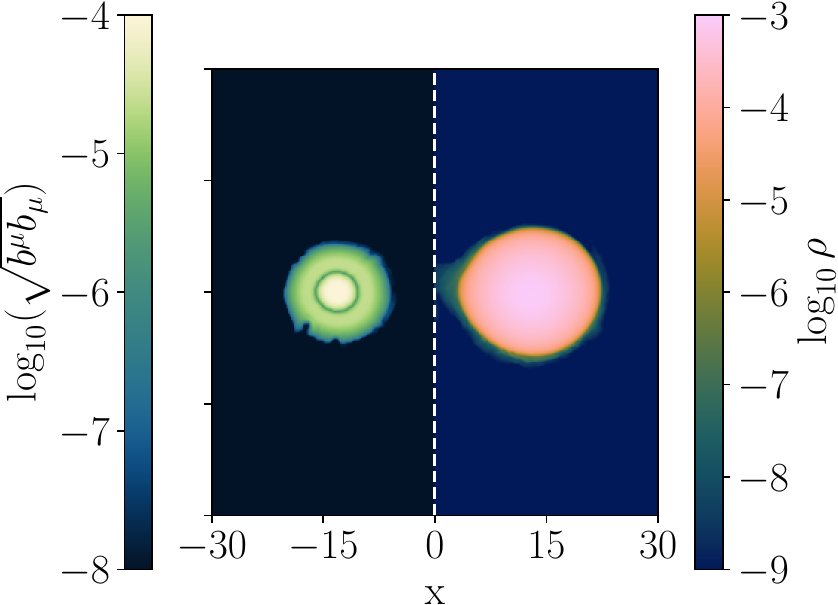}
      \label{fig:hybrid_bns_inspiral_igm2}
    }
    &
    \subfloat[$t=944$ (two full orbits).]{
      % \addtocounter{subfigure}{2}
      \includegraphics[width=0.45\textwidth,clip]{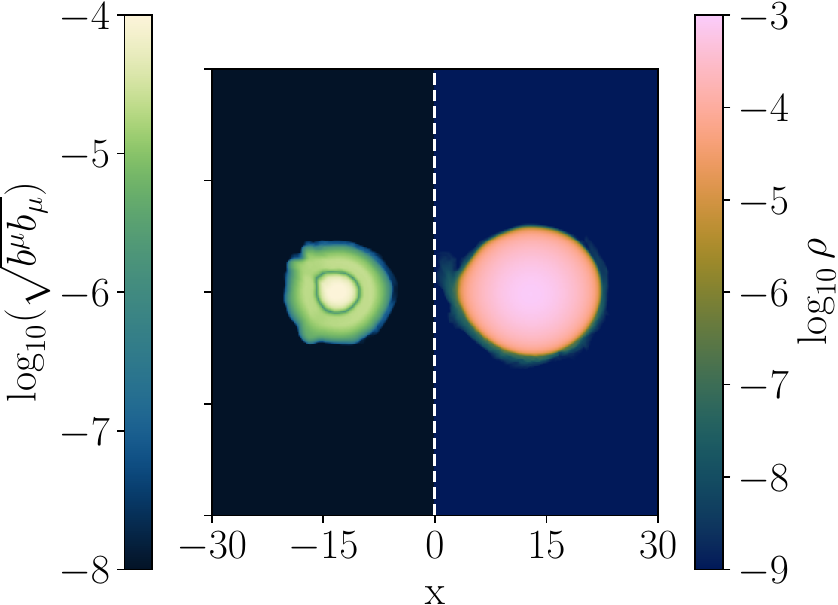}
      \label{fig:hybrid_bns_inspiral_grhayl2}
    }
    \\
    \subfloat[$t=1632$ (first touch).]{
      % \addtocounter{subfigure}{-3}
      \includegraphics[width=0.45\textwidth,clip]{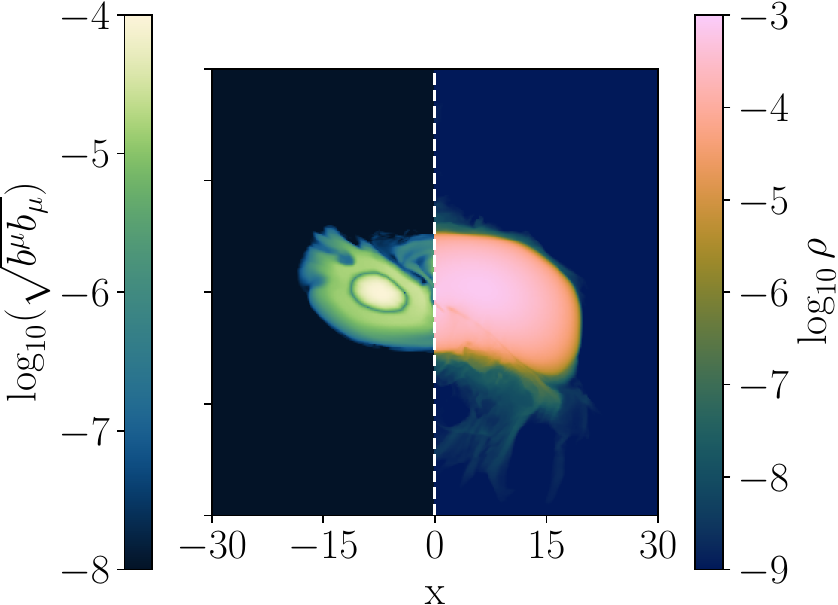}
      \label{fig:hybrid_bns_inspiral_igm3}
    }
    &
    \subfloat[$t=1632$ (first touch).]{
      % \addtocounter{subfigure}{2}
      \includegraphics[width=0.45\textwidth,clip]{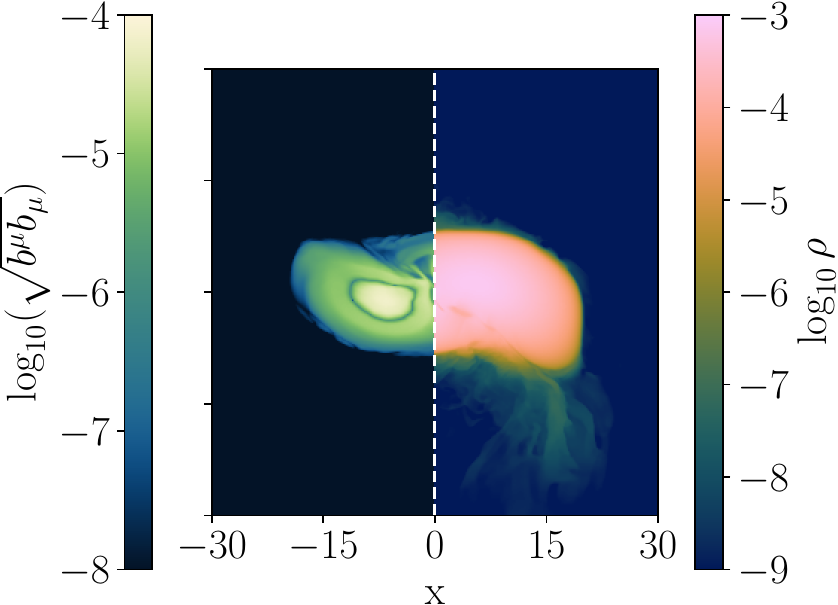}
      \label{fig:hybrid_bns_inspiral_grhayl3}
    }
  \end{tabular}
  \caption{Log plots of data from a magnetized, equal-mass BNS during inspiral with a hybrid \eos. The left side shows the $xy$-plane for \origigm, and the right side shows the same for \igm. For all panels, the right half shows the density, and the left half shows the magnetic field strength $\sqrt{b^\mu b_\mu}$.}
\label{fig:hybrid_bns_splot_inspiral}
\end{figure*}

\begin{figure*}[p]
  \centering
  \setlength{\abovecaptionskip}{10pt}
  \begin{tabular}{c|c}
    \origigm & \igm \\
    \subfloat[$t=1920$ (hypermassive neutron star).]{
      \includegraphics[width=0.45\textwidth,clip]{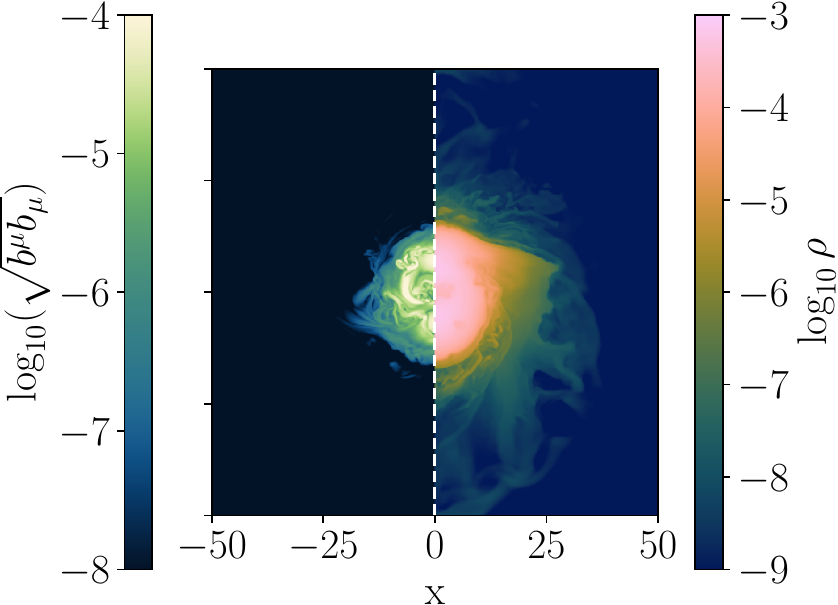}
      \label{fig:hybrid_bns_hns_igm1}
    }
    &
    \subfloat[$t=1920$ (hypermassive neutron star).]{
      % \addtocounter{subfigure}{2}
      \includegraphics[width=0.45\textwidth,clip]{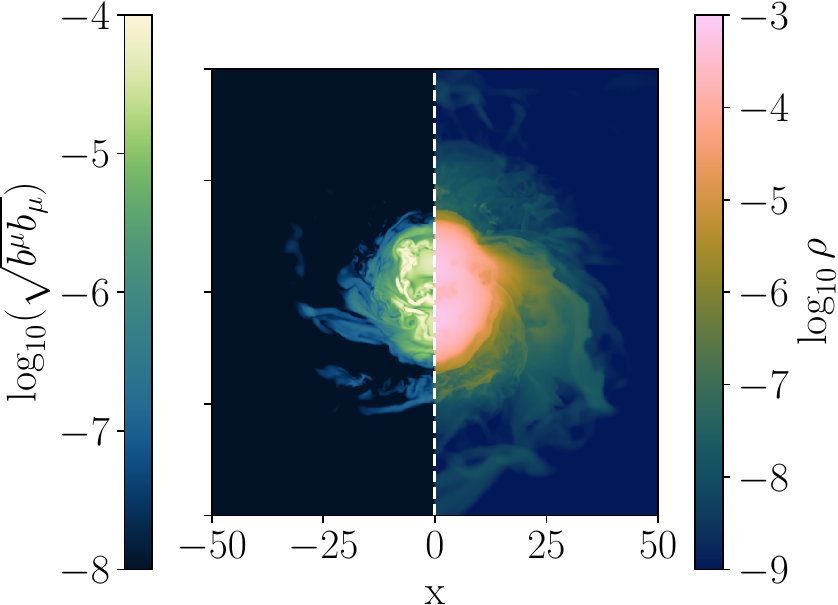}
      \label{fig:hybrid_bns_hns_grhayl1}
    }
    \\
    \subfloat[$t=2768$ (half-way to collapse).]{
      % \addtocounter{subfigure}{-3}
      \includegraphics[width=0.45\textwidth,clip]{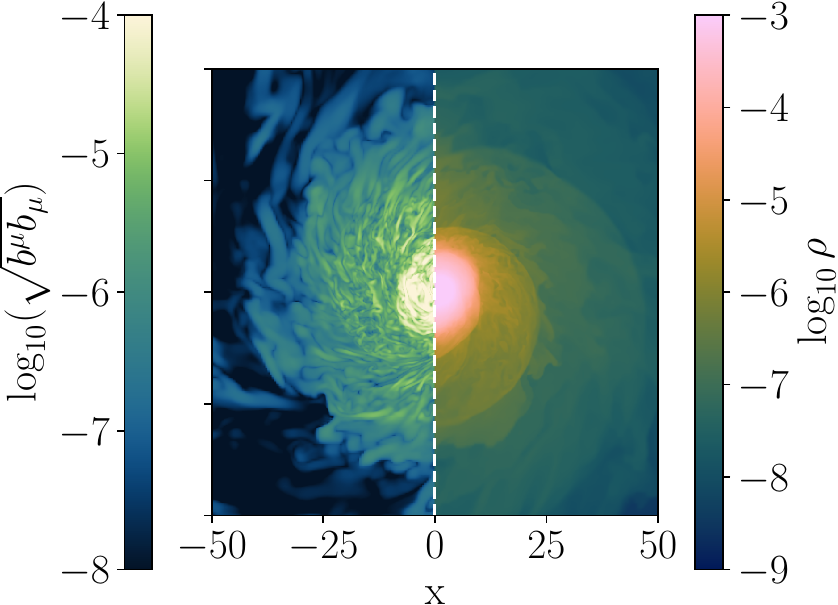}
      \label{fig:hybrid_bns_hns_igm2}
    }
    &
    \subfloat[$t=2768$.]{
      % \addtocounter{subfigure}{2}
      \includegraphics[width=0.45\textwidth,clip]{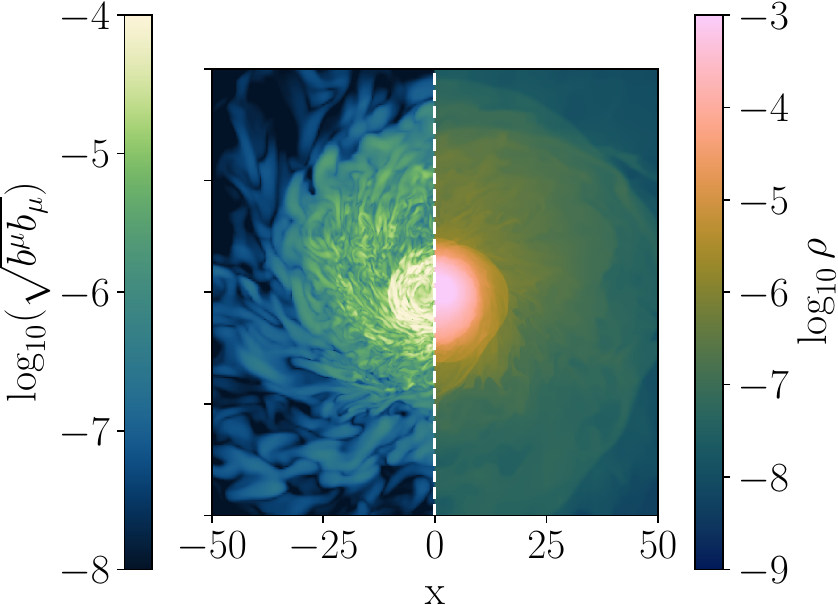}
      \label{fig:hybrid_bns_hns_grhayl2}
    }
    \\
    \subfloat[$t=3600$ (postcollapse).]{
      % \addtocounter{subfigure}{-3}
      \includegraphics[width=0.45\textwidth,clip]{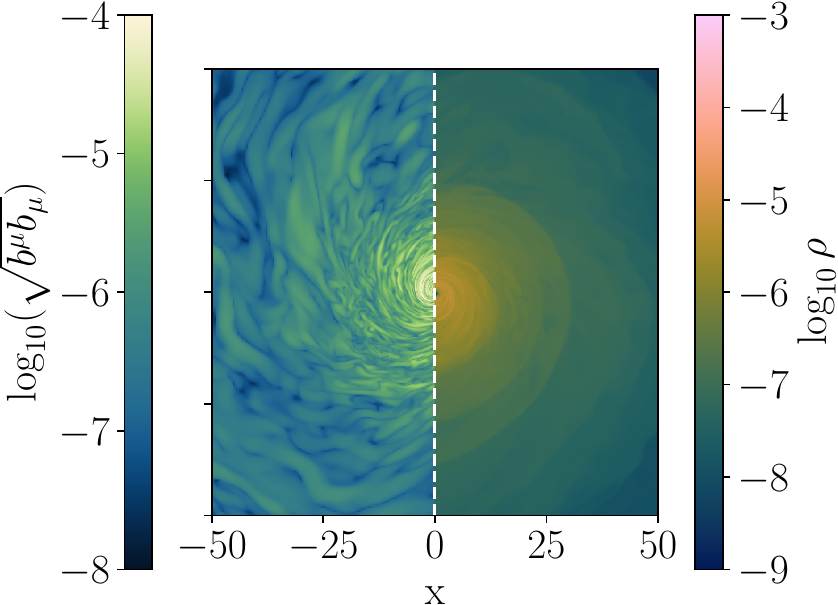}
      \label{fig:hybrid_bns_hns_igm3}
    }
    &
    \subfloat[$t=4640$ (postcollapse).]{
      % \addtocounter{subfigure}{2}
      \includegraphics[width=0.45\textwidth,clip]{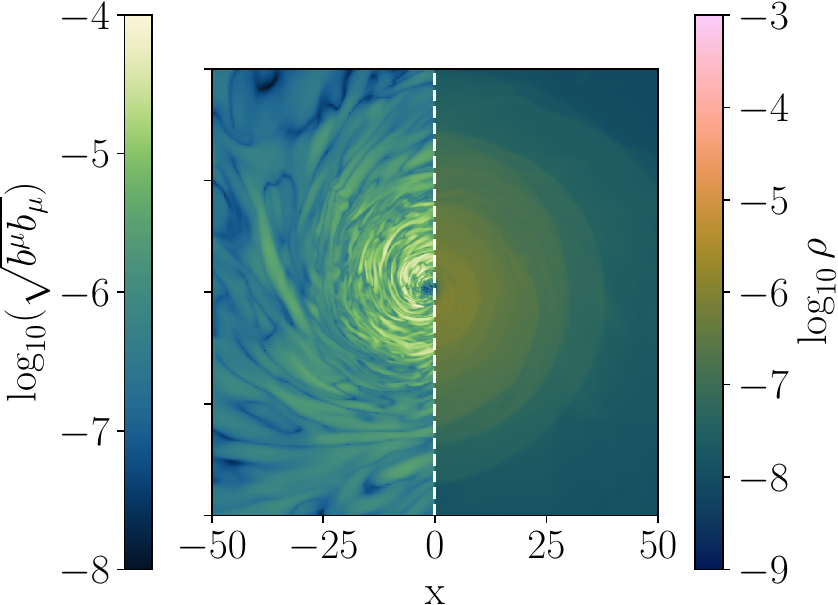}
      \label{fig:hybrid_bns_hns_grhayl3}
    }
  \end{tabular}
  \caption{Same as \figref{fig:hybrid_bns_splot_inspiral} for the postmerger phase.}
\label{fig:hybrid_bns_splot_hns}
\end{figure*}
%\clearpage

\begin{figure*}[ht!]
  \centering
	\subfloat{\includegraphics[width=0.48\linewidth]{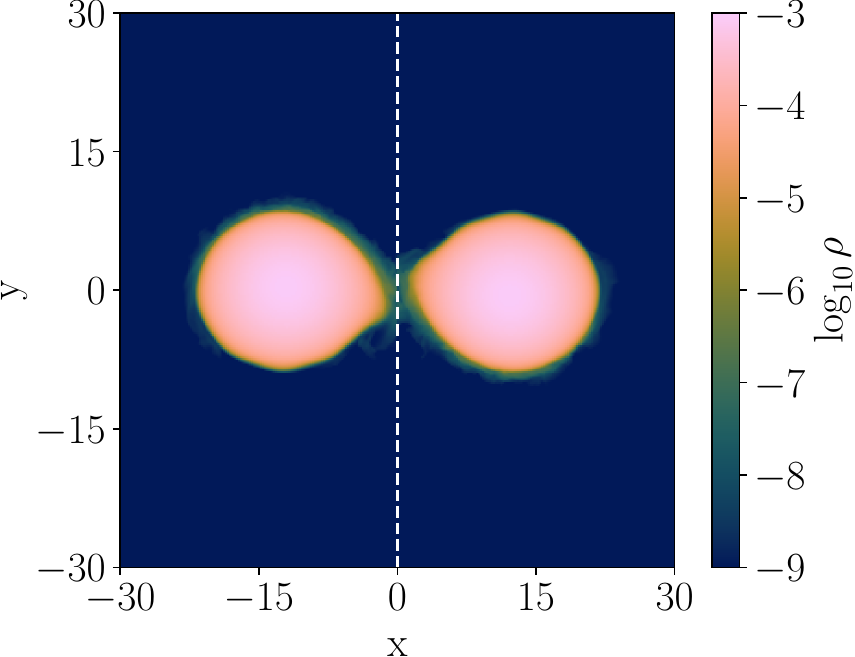}}
    \hfill
	\subfloat{\includegraphics[width=0.48\linewidth]{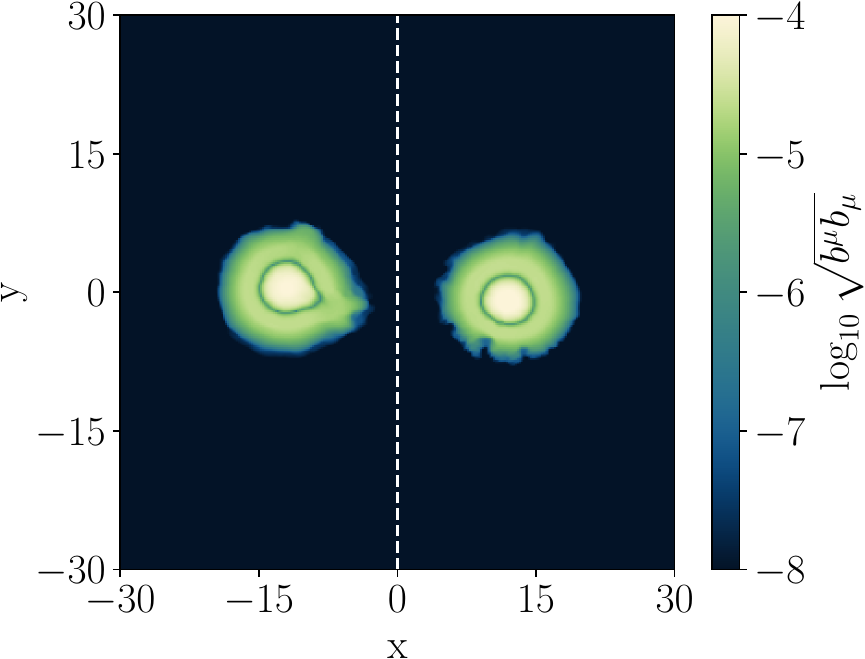}}
    \caption{Log plots of density (\textbf{left}) and magnetic field strength (\textbf{right}) at $t=1216$. Both plots are split with data from \igm (\textbf{left}) and \origigm (\textbf{right}).}
    \label{fig:hybrid_surface1}
\end{figure*}

\begin{figure*}[ht!]
  \centering
    \subfloat{\includegraphics[width=0.48\linewidth]{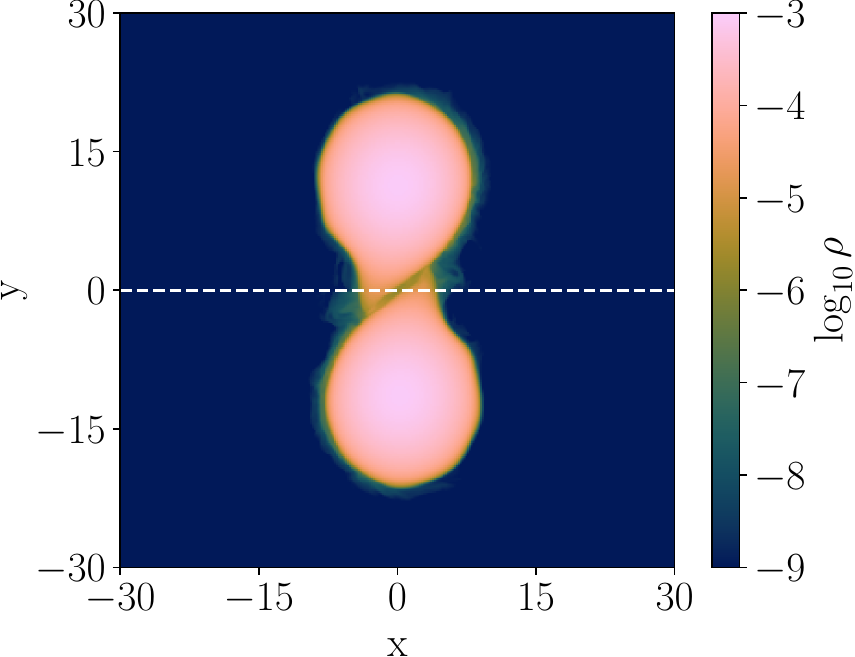}}
    \hfill
    \subfloat{\includegraphics[width=0.48\linewidth]{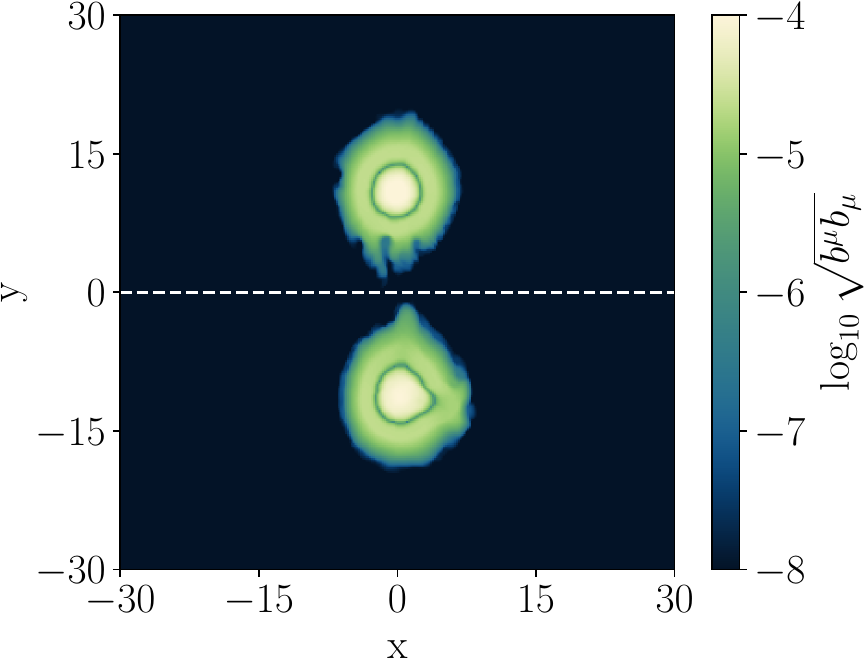}}
    \caption{Log plots of density (\textbf{left}) and magnetic field strength (\textbf{right}) at $t=1344$. Both plots are split with data from \igm (\textbf{bottom}) and \origigm (\textbf{top}).}
    \label{fig:hybrid_surface2}
\end{figure*}

Of course, simply tracking the central density does not fully assess the physics of the simulation. To further examine the results, we show spatial slices of density and magnetic field strength $\sqrt{b^\mu b_\mu}$ for both codes. In \figref{fig:hybrid_bns_splot_inspiral}, we show the inspiral at three different times. As expected, the initial data in \figref{fig:hybrid_bns_inspiral_igm1} and \figref{fig:hybrid_bns_inspiral_grhayl1} match exactly. While it is not as clear with still images, the distortions that appear in the magnetic field strength during the inspiral are noticeably worse with \origigm. The stellar surface in \figref{fig:hybrid_bns_inspiral_igm2} and \figref{fig:hybrid_bns_inspiral_igm3} shows far worse behavior than in \figref{fig:hybrid_bns_inspiral_grhayl2} and \figref{fig:hybrid_bns_inspiral_grhayl3}. The chipped appearance of the isocontours is completely absent in the new code. While it develops some asymmetric aspects during the inspiral, the magnetic fields are advected better than in the original code.

In \figref{fig:hybrid_bns_splot_hns}, we show three spatial slices for the postmerger. Note that the collapse time is different between the codes. The first two sets of plots are for the same time and display very similar behavior. The magnetic field strength again follows the matter flows better in \igm, with the matter tails coming off the merging binary having accompanying magnetic components that are missing in \origigm. Figures~\ref{fig:hybrid_bns_hns_igm3} and \ref{fig:hybrid_bns_hns_grhayl3} show the postcollapse system after the central mass has settled down into a black hole with inspiraling matter. Both codes show very similar dynamics and remain stable out to $t=6000$.

To illustrate the differences more clearly, Figures~\ref{fig:hybrid_surface1} and \ref{fig:hybrid_surface2} show the density and magnetic field strength for both codes. \origigm has very noticeable distortions in magnetic field topology near the star surface, while \igm has a much smoother profile. Further, the magnetic field in \igm follows the distortion of the stars, with the deformation leading to merger causing a deformation of the magnetic field strength, as seen in \figref{fig:hybrid_surface1}. In contrast, \origigm remains nearly spherical.

As the stars enter the merger, the inspiral rate outpaces the magnetic field's ability to keep up, leading to the bulge in magnetic field strength visible in \figref{fig:hybrid_surface2}. The deformation toward positive $x$ for the lower star is the same deformation present in \figref{fig:hybrid_surface1}, as this feature fails to follow the star's rotation through the last quarter rotation into the merger. This is not apparent in \origigm since it remains nearly spherical all the way to merger.

\subsubsection{Hybrid \eos, Piecewise Polytrope}

The second BNS test we perform adopts the piecewise polytropic representation of the SLy \eos from~\cite{Read:2008iy}. We use \code{FUKA}~\cite{FUKA} to generate initial data for an equal-mass system, with each star having a mass of $1.4\Msun$ and radius ${\approx}\qty{8.86}{\km}$. We seed the interior of each neutron star with a poloidal magnetic field with maximum initial value ${\sim}10^{15}~G$ (see Appendix C of~\cite{Etienne:2015cea}).

We set up a grid with eight refinement levels by factors of 2, with the resolution at the finest level \mbox{$\dx_{8}\approx185~\mathrm{m}$}. After black hole (BH) formation, two additional refinement levels are added to resolve the puncture, making the highest grid resolution \mbox{$\dx_{10}\approx46~\mathrm{m}$}.

\begin{figure}[ht!]
  \centering
  \includegraphics[width=\columnwidth]{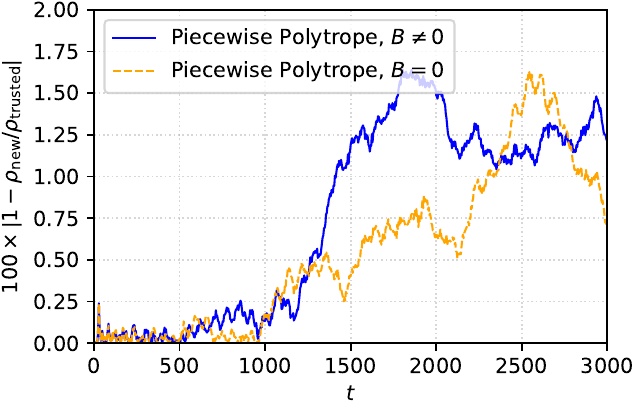}
  \caption{Relative error in maximum density over time between \igm and \origigm for BNS simulations with (blue, solid) and without (orange, dashed) magnetic fields.}
  \label{fig:ppeos_bns_central_density}
\end{figure}

In \figref{fig:ppeos_bns_central_density}, we display a time series of the relative difference in the maximum baryonic density over the numerical grid between \igm and \origigm. The plot is cut off at $t=3000$, just prior to BH formation, after which the diagnostic becomes unreliable. Over the displayed time period, we see that the two PP\eos implementations agree to \qty{2.5}{\%}.

\begin{figure*}[p]
  \def\FigWidth{0.375\textwidth}
  \def\FigSpace{-2.5ex}
  \centering
  \setlength{\abovecaptionskip}{10pt}
  \begin{tabular}{c|c}
    \origigm & \igm \\
    \subfloat[$t=0$.]{
      \includegraphics[width=\FigWidth,clip]{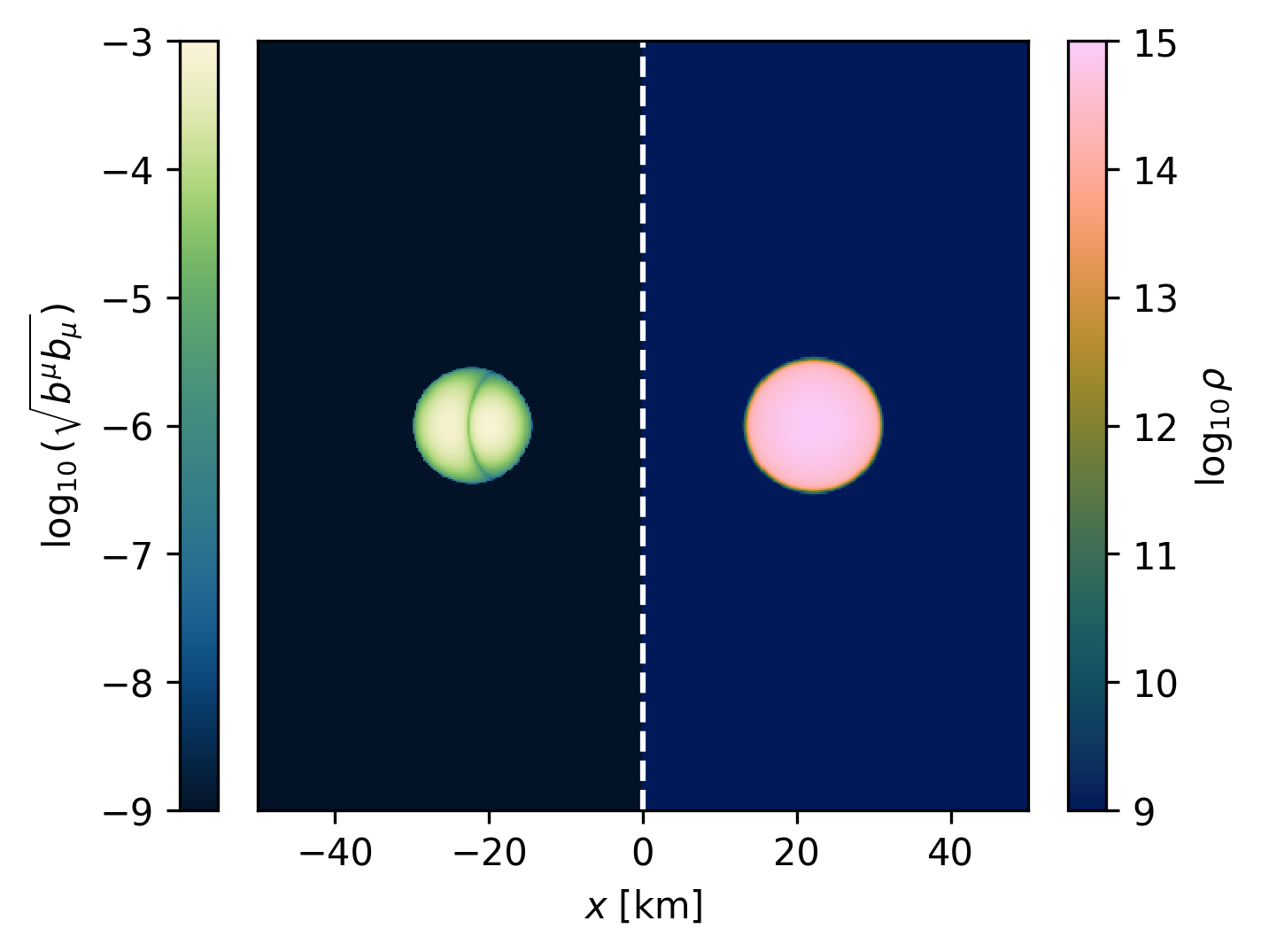}
      \label{fig:ppeos_bns_igm_initial}
    }
    &
    \subfloat[$t=0$.]{
      % \addtocounter{subfigure}{2}
      \includegraphics[width=\FigWidth,clip]{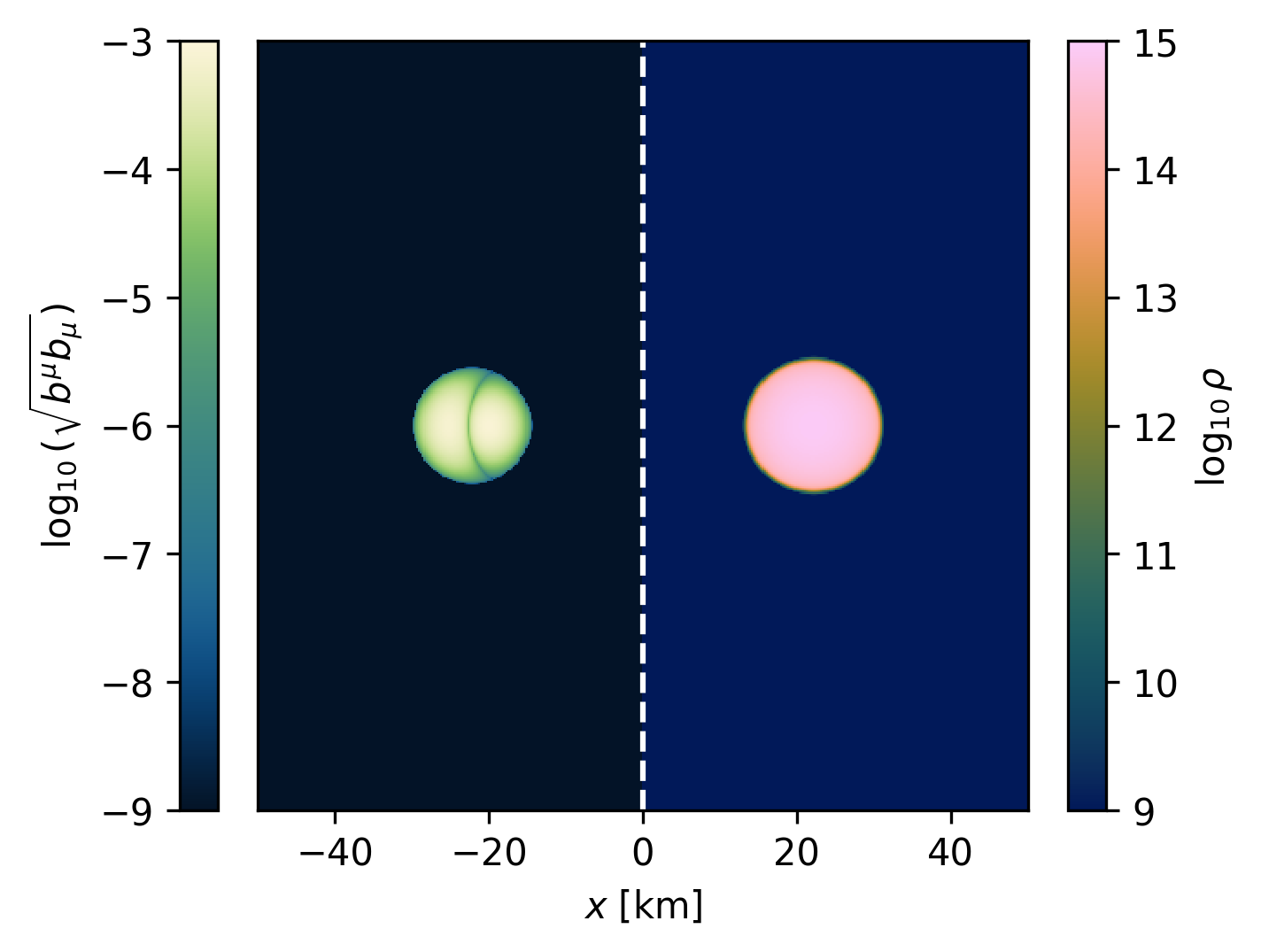}
      \label{fig:ppeos_bns_ghl_initial}
    }
    \\[\FigSpace]
    \subfloat[$t=\qty{8.51}{\ms}$ (${\approx}2.5$ orbits).]{
      % \addtocounter{subfigure}{-3}
      \includegraphics[width=\FigWidth,clip]{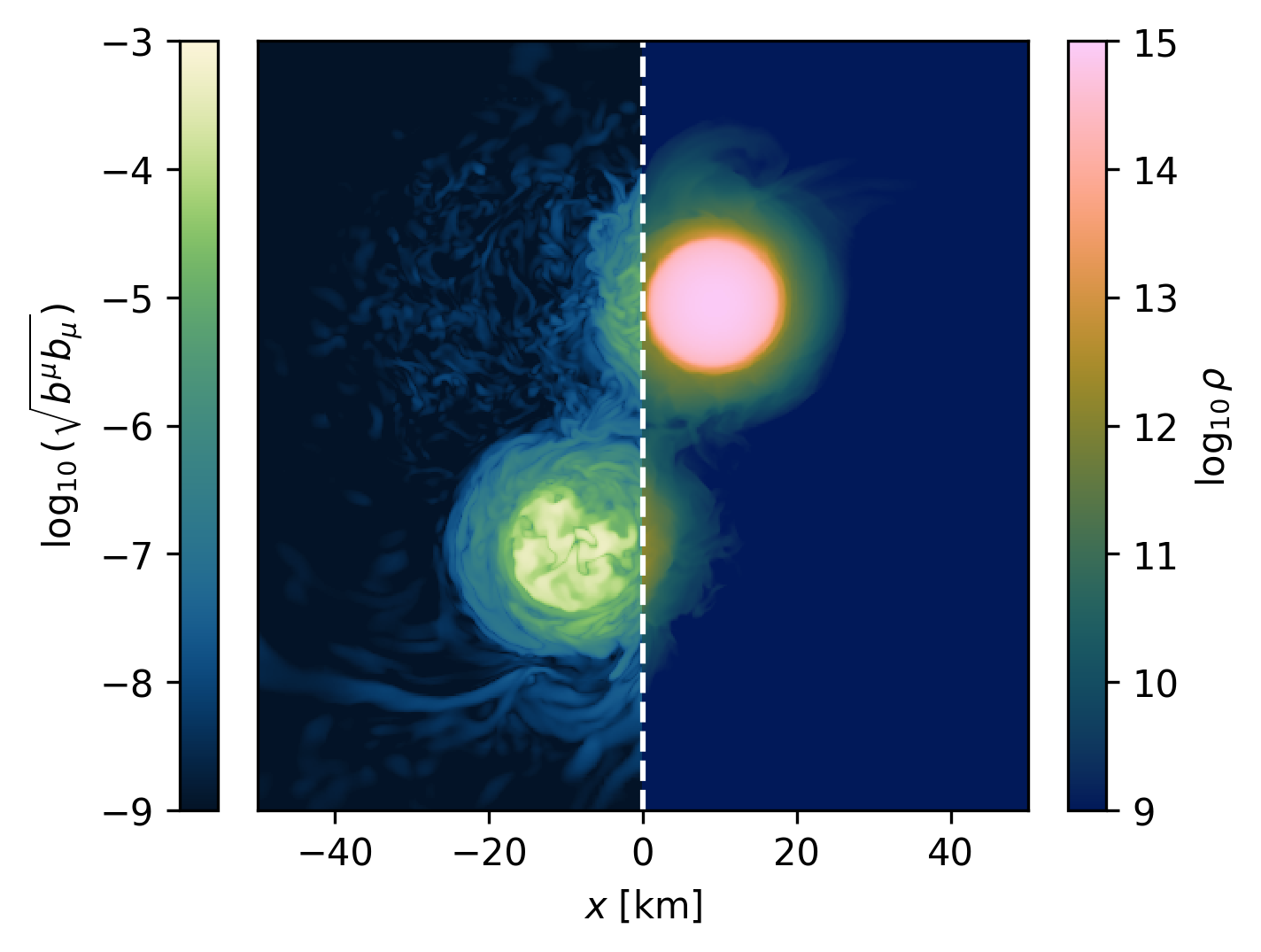}
      \label{fig:ppeos_bns_igm_2.5ish}
    }
    &
    \subfloat[$t=\qty{8.51}{\ms}$ (${\approx}2.5$ orbits).]{
      % \addtocounter{subfigure}{2}
      \includegraphics[width=\FigWidth,clip]{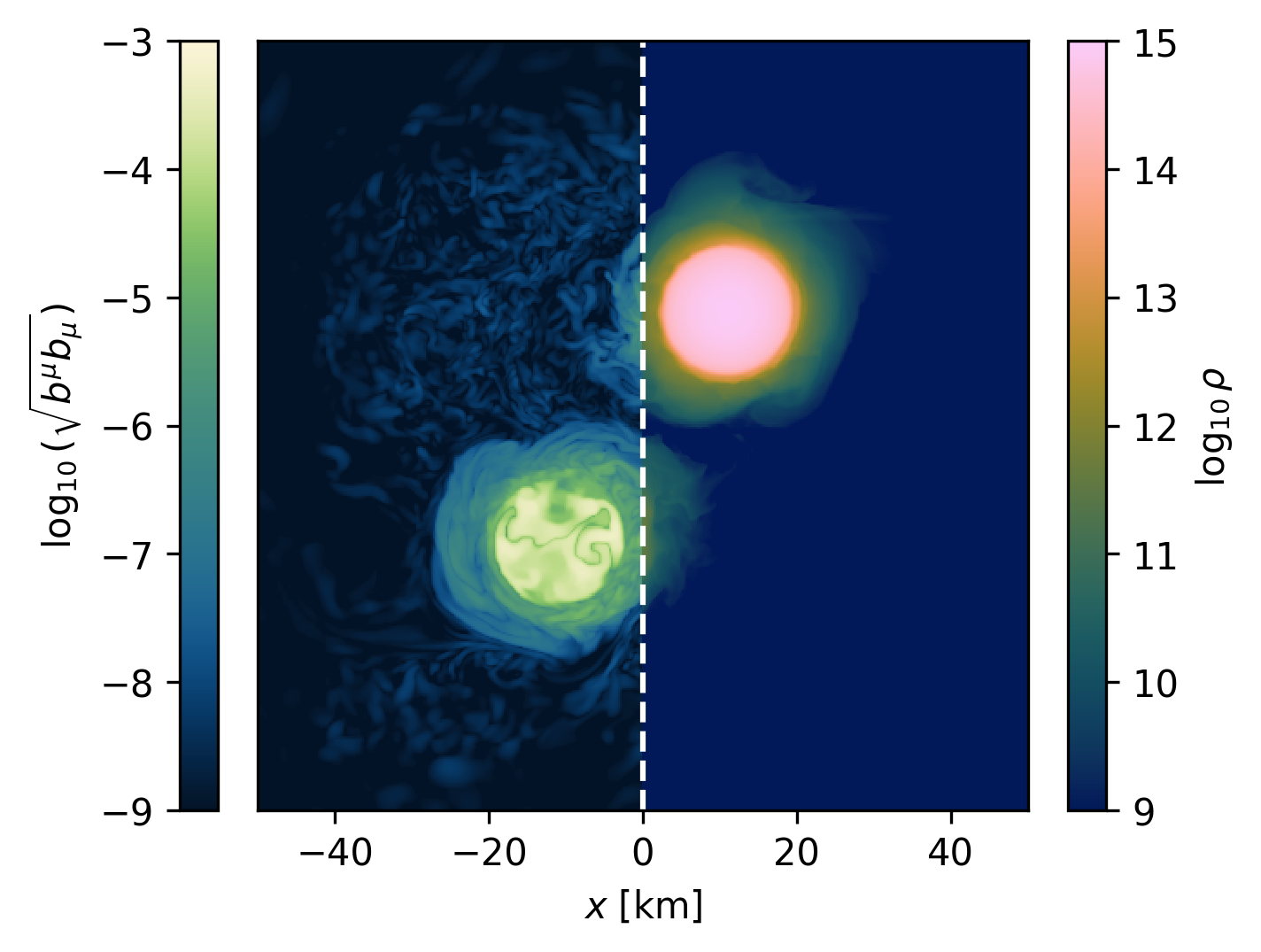}
      \label{fig:ppeos_bns_ghl_2.5ish}
    }
    \\[\FigSpace]
    \subfloat[$t=\qty{15.3}{\ms}$ (first touch).]{
      % \addtocounter{subfigure}{-3}
      \includegraphics[width=\FigWidth,clip]{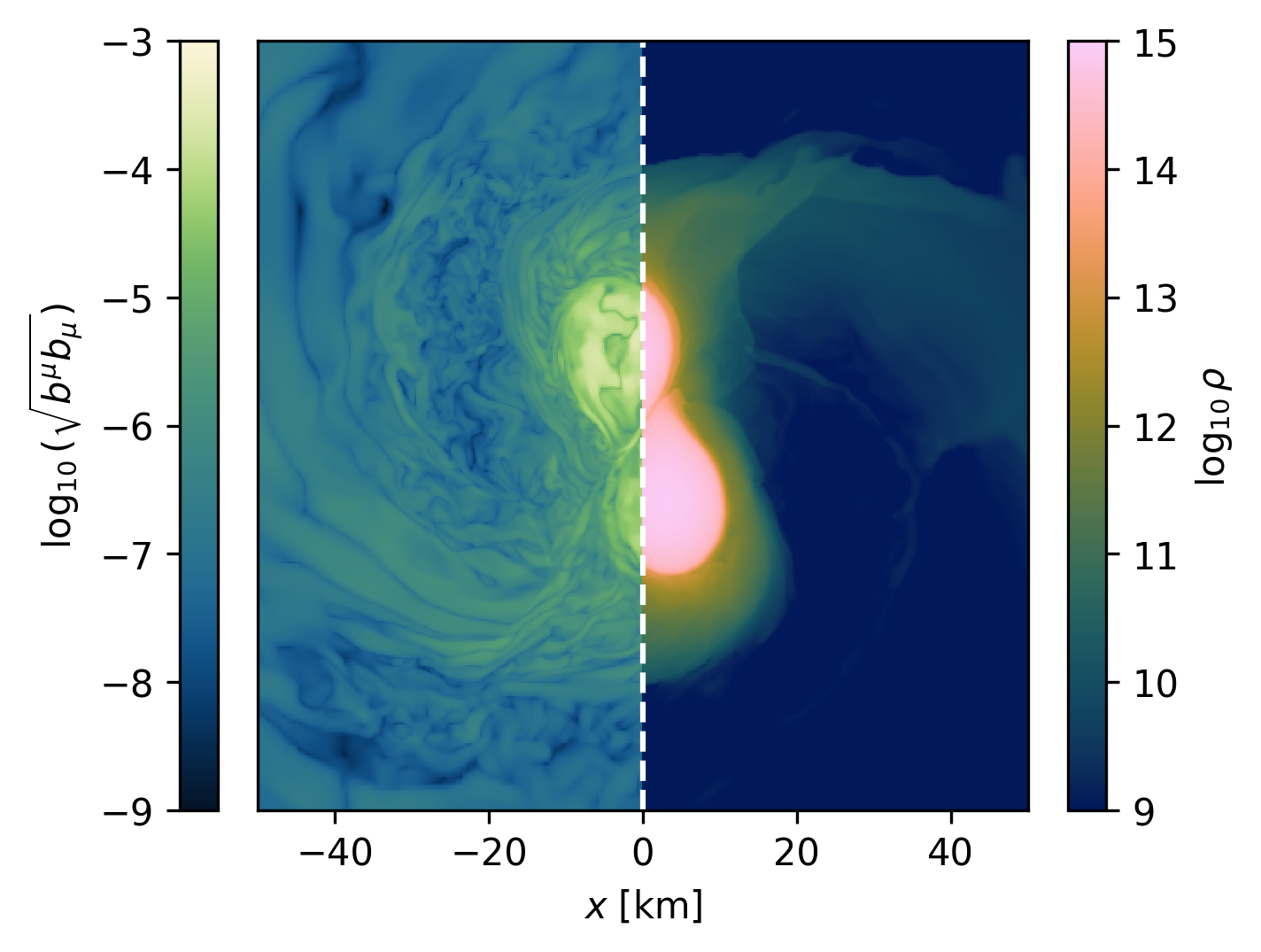}
      \label{fig:ppeos_bns_igm_touch}
    }
    &
    \subfloat[$t=\qty{15.4}{\ms}$ (first touch).]{
      % \addtocounter{subfigure}{2}
      \includegraphics[width=\FigWidth,clip]{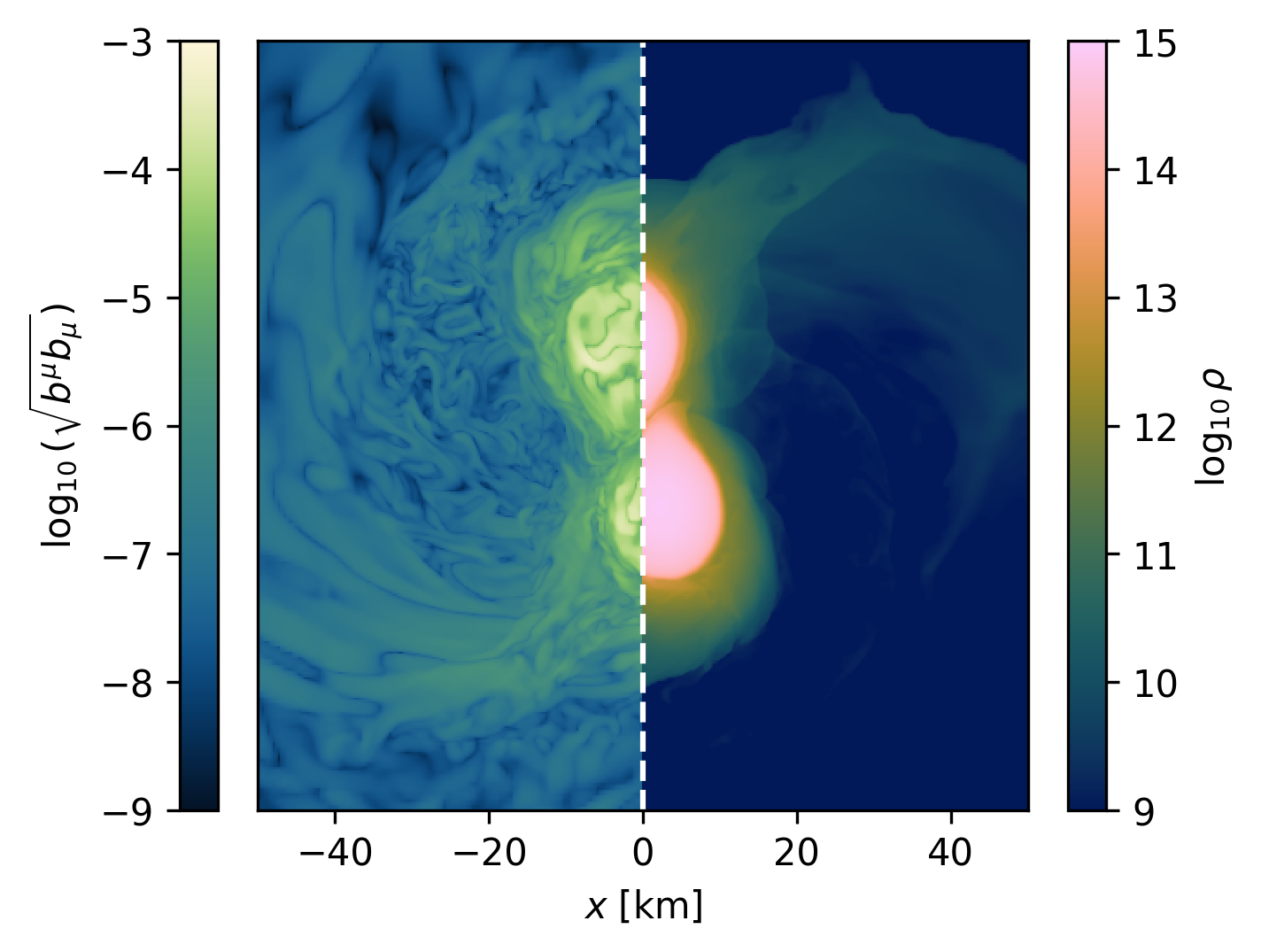}
      \label{fig:ppeos_bns_ghl_touch}
    }
    \\[\FigSpace]
    \subfloat[$t=\qty{17.2}{\ms}$ (moments after BH formation).]{
      % \addtocounter{subfigure}{-3}
      \includegraphics[width=\FigWidth,clip]{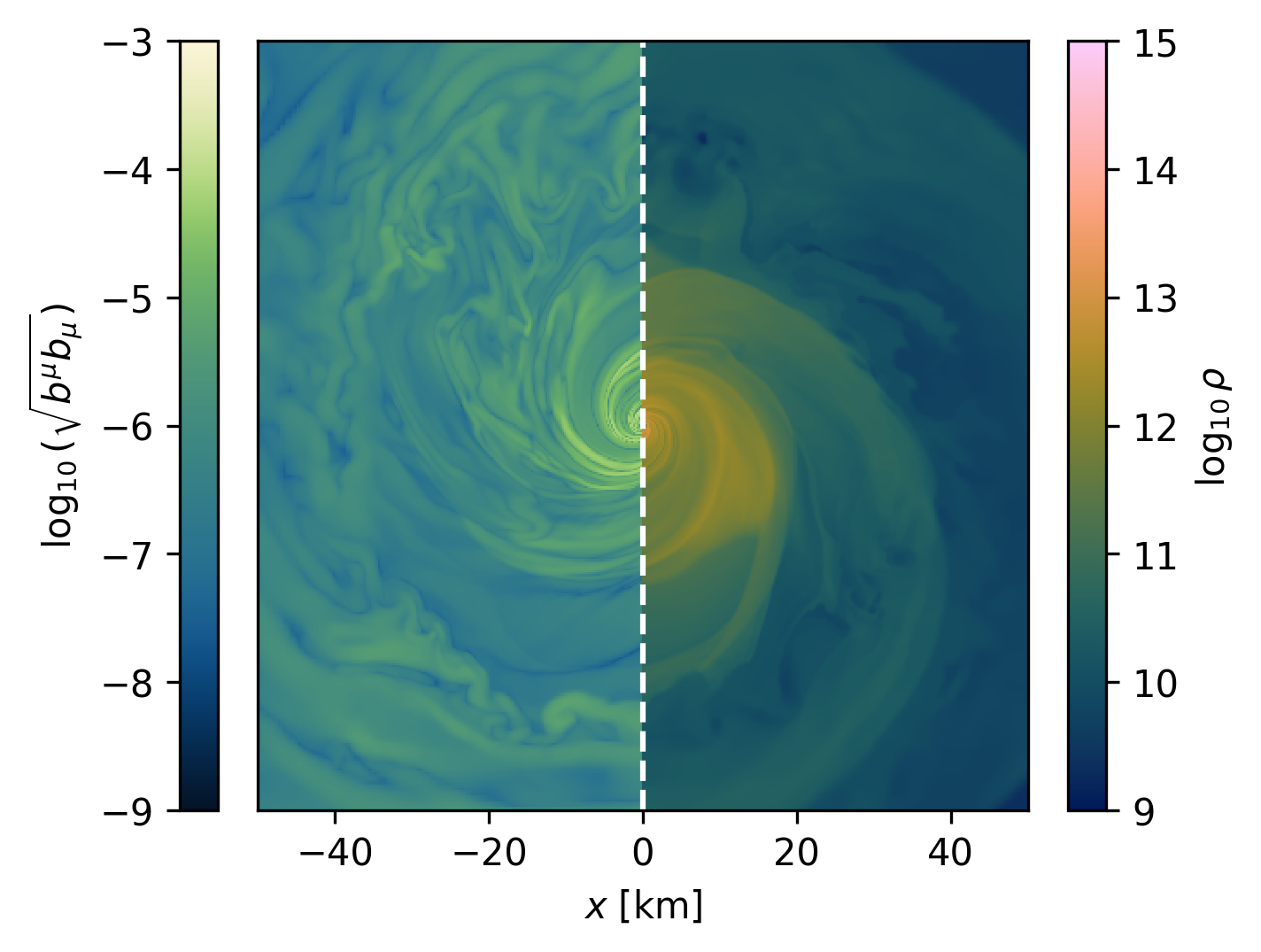}
      \label{fig:ppeos_bns_igm_bh}
    }
    &
    \subfloat[$t=\qty{17.2}{\ms}$ (moments after BH formation).]{
      % \addtocounter{subfigure}{2}
      \includegraphics[width=\FigWidth,clip]{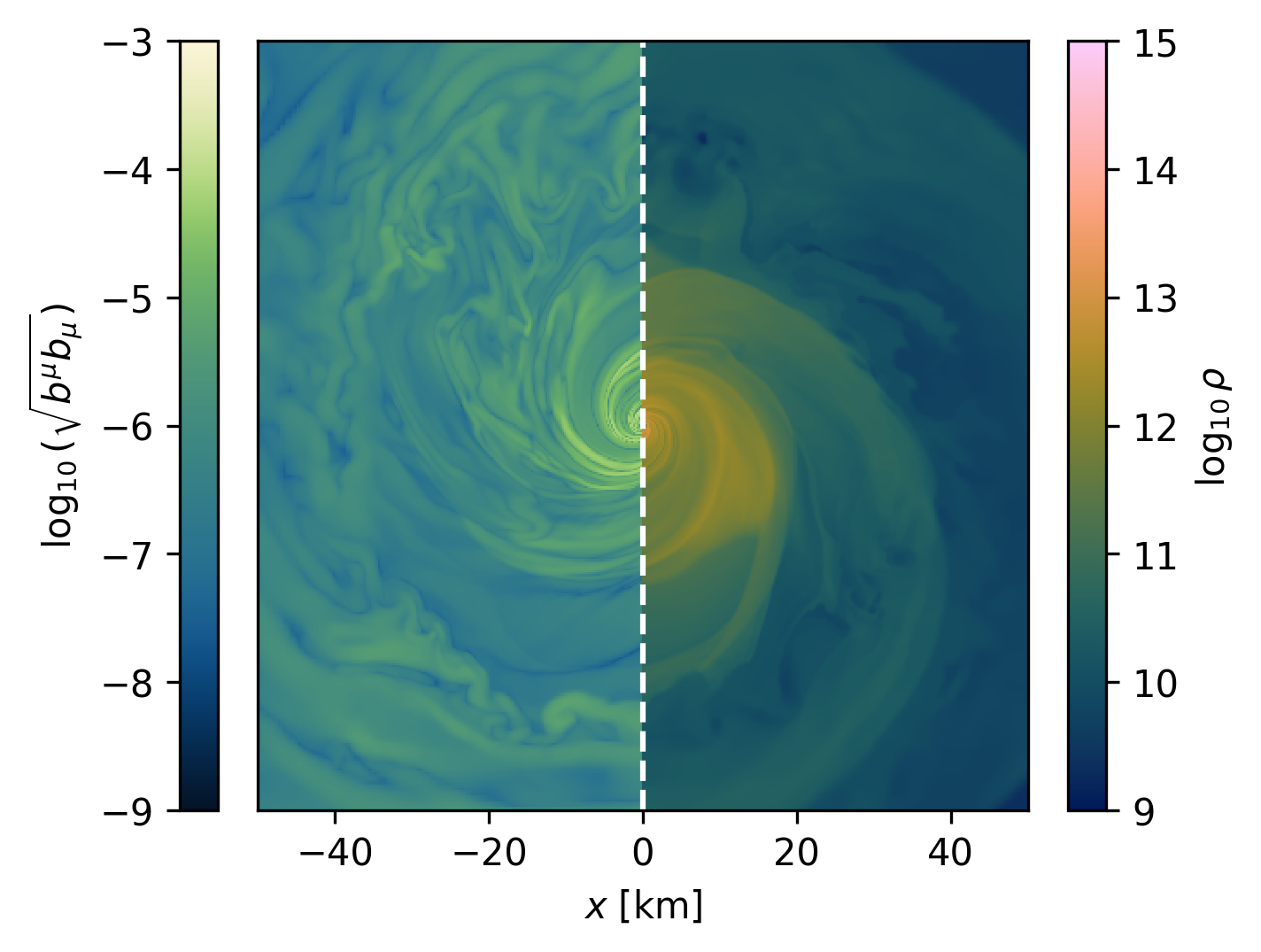}
      \label{fig:ppeos_bns_ghl_bh}
    }
  \end{tabular}
  \caption{Baryonic density (in cgs) and magnetic pressure of an equal-mass, magnetized BNS simulation. We can observe a slight dephasing between \igm and \origigm, resulting in a ${\approx}\qty{0.1}{\ms}$ difference in the time it takes for the stars to first touch, followed by prompt collapse to a BH.}
  \label{fig:ppeos_bns_inspiral}
\end{figure*}

Figure~\ref{fig:ppeos_bns_inspiral} displays several snapshots of the baryonic density and magnetic pressure $P_{\rm mag} = b^{2}/2$ at different stages of the simulation. We note the slight dephasing between \igm and \origigm, which results in the stars first touching ${\approx}\qty{0.1}{\ms}$ apart. BH formation occurs promptly after. We again emphasize the similarity between the results, and draw the reader's attention to how \igm better advects magnetic fields, just like in the case of a simple polytropic \eos.

\section{Concluding remarks}
\label{sec:concluding_remarks}

\grhayl provides an effective method for quickly developing GRMHD evolution code in new infrastructures. Any changes to the library are validated by continuous-integration testing (see Appendix~\ref{app:CI}). Few GRMHD codes perform regular testing at the level of individual functions (and demand round-off-level agreement), and these tests represent professional industry standards that are rarely implemented in open-source scientific codes.

The current library contains all the necessary pieces to quickly implement an \igm-like code in new infrastructures. This includes multiple \eos types, Con2Prim and reconstruction methods, hydrodynamic and induction equation RHSs, and neutrino leakage. In addition to the current features, plans include adding more advanced radiation transport schemes, more atmosphere prescriptions and more Con2Prim options in the near future. Work is already underway to extend the Noble implementations to support tabulated \eos, as well as adding the 2D and 3D Cerd\'a-Dur\'an~\etal routines~\cite{Cerda_2008}. Support for M1 closure neutrino transport based on~\cite{Radice:2021jtw} is also planned.

Implementation of \bhah's multipatch, multicoordinate infrastructure within \groovy will enable highly efficient BNS simulations. Work also continues toward adding magnetic evolution to the curvilinear code, which must be handled with particular care due to the combination of staggered grids and coordinate singularities.

Several more GRMHD implementations using \grhayl are already in the planning or development phase for other infrastructures, some of which currently lack hydrodynamics codes. While the community benefits from multiple codes for validation and scientific rigor, the advantages of having a code that can easily be integrated into any infrastructure are significant. These advantages include easier vetting of new infrastructures, faster implementation of GRMHD code in future HPC environments, and establishing a cross-infrastructure ``standard'' that can more easily determine whether differences between codes arise from the physics code or the underlying infrastructure.

\section*{Acknowledgments}
We thank Erik Schnetter and Roland Haas for their help with using the new \carpetx driver. We also thank Jay Kalinani for discussions about the \code{AsterX} GRMHD code which also uses \carpetx, as well as discussions regarding implementing the simple \eos.

T.P.J.\ gratefully acknowledges funding support from NASA FINESST-80NSSC23K1437, West Virginia University's Chancellor's Scholarship program, and the Southern Regional Education Board's Dissertation Award program. Z.B.E.\ gratefully acknowledges funding support from NSF Awards No.\ AST-2227080, PHY-2110352, OAC-2227105, and NASA Awards No.\ ISFM-80NSSC21K1179 and TCAN-80NSSC18K1488.

This research made use of the resources of the High Performance Computing Center at Idaho National Laboratory, which is supported by the Office of Nuclear Energy of the U.S. Department of Energy and the Nuclear Science User Facilities under Contract No. DE-AC07-05ID14517.

\section*{Data Availability}
The data that support the findings of this article are not publicly available upon publication because it is not technically feasible and/or the cost of preparing, depositing, and hosting the data would be prohibitive within the terms of this research project. The data are available from the authors upon reasonable request.

\appendix

\section{Basic Equations}
\label{sec:basic_equations}

For completeness, we provide a brief overview of the GRMHD equations and choice of variables used by \grhayl and the core equations that it is designed to solve. As is standard in a numerical relativity framework, we use the 3+1 formalism, in which the spacetime metric is written as
\begin{equation}
    ds^{2} = \left(-\alpha^{2}+\beta^{i}\beta_{i}\right)dt^{2}
        + 2\beta_{i}\,dt\, dx^{i}
        + \gamma_{ij}\,dx^{i}\,dx^{j}\;,
\end{equation}
where $\alpha$ is the lapse function, $\beta^{i}$ is the shift vector, and $\gamma_{ij}$ is the ADM physical spatial metric.

The GRMHD equations---conservation of baryon number, conservation of lepton number, conservation of energy-momentum, and homogeneous Maxwell's equations---are given by
\begin{align}
    \nabla_{\mu}\left(\numb u^{\mu}\right) &= 0\;,\label{eq:baryon_number_conservation}\\
    \nabla_{\mu}\left(\nume u^{\mu}\right) &= 0\;,\label{eq:lepton_number_conservation}\\
    \nabla_{\mu}T^{\mu\nu} &= 0\;,\label{eq:enmom_conservation}\\
    \nabla_{\mu}\Fdual^{\mu\nu} &= 0\;,\label{eq:maxwell}
\end{align}
respectively. Here, $\numb$ and $\nume$ are the number densities of baryons and leptons, \mbox{$\rho = \numb\mb$} the baryon density, $\mb$ the baryon mass, and \mbox{$\ye=\nume/\numb$} the electron fraction. Further, $u^{\mu}$ is the fluid four-velocity, \mbox{$\Fdual^{\mu\nu}=(1/2)\tilde{\epsilon}^{\mu\nu\rho\sigma}F_{\rho\sigma}$} is the dual of the Faraday tensor $F^{\mu\nu}$, and $\tilde{\epsilon}^{\mu\nu\rho\sigma}$ is the Levi-Civita tensor. Finally, we assume ideal MHD (\mbox{$u_{\mu}F^{\mu\nu}=0$}) throughout.

The evolution equations are written in flux-conservative form via
\begin{equation}
    \partial_{t} \consv + \nabla_{j}\fluxv^{j} = \sourcev\;,
    \label{eq:basic_evol}
\end{equation}
where $\consv$ is the vector of conservative variables, $\fluxv^{j}$ is the flux vector along direction $j$, and $\sourcev$ is the vector of source terms. The vectors $\consv$, $\fluxv^{j}$, and $\sourcev$ depend directly on the so-called primitive variables $\primv$, defined as
\begin{equation}
\primv = \begin{bmatrix}
    \rho\\
    \ye\\
    S\\
    P\\
    v^{i}\\
    B^{i}
\end{bmatrix}\;.
\label{eq:prims}
\end{equation}
where $P$ is the pressure, \mbox{$v^{i} \equiv u^{i}/u^{0}$} is the fluid three-velocity, and $S$ is the entropy. Note that this velocity is different from the velocity used in the Valencia formulation~\cite{Banyuls_1997, Marti_1991}
\begin{equation}
  \tilde{u}^i = \alpha^{-1}\bigl(v^{i} + \beta^{i}\bigr)\;,
  \label{eq:valencia_v}
\end{equation}
which is measured by Eulerian observers with unit four-velocity \mbox{$n^{\mu} = \left(\alpha^{-1},\alpha^{-1}\beta^{i}\right)$} normal to the spatial hypersurfaces.

Similarly, there are two common definitions of $B^{\mu}$, the magnetic field measured by Eulerian observers.
% with unit four-velocity \mbox{$n^{\mu} = \left(\alpha^{-1},\alpha^{-1}\beta^{i}\right)$} normal to the spatial hypersurfaces.
The first, adopted by e.g.,~\cite{Duez_IGM}, is
\begin{equation}
    B^\mu = n_\nu \Fdual^{\mu\nu}\;,
    \label{eq:Bdef1}
\end{equation}
while the second, adopted by e.g.,~\cite{GRHydro}, is
\begin{equation}
    B^\mu = \frac{1}{\sqrt{4\pi}} n_\nu \Fdual^{\mu\nu}\;.
    \label{eq:Bdef2}
\end{equation}
Unlike the original version of \igm, which adopts \eqref{eq:Bdef1}, \grhayl adopts \eqref{eq:Bdef2}. This choice does not affect the evolution scheme in any way, but it is the more natural choice from a numerical perspective, avoiding unnecessary multiplications and divisions by factors of $\sqrt{4\pi}$.

The conservative variables $\consv$ (minus the magnetic field) in \eqref{eq:basic_evol} are given by
\begin{equation}
    \consv =
    \begin{bmatrix}
        \rho_*\\
        \yet\\
        \tilde{S}\\
        \tilde{\tau}\\
        \tilde{S}_{i}
    \end{bmatrix}
    \equiv
    \sqrtgamma
    \begin{bmatrix}
        W\rho\\
        W\rho\ye\\
        W S\\
        \tau\\
        S_{i}
    \end{bmatrix}
    =
    \sqrtgamma
    \begin{bmatrix}
        W\rho\\
        W\rho\ye\\
        W S\\
        \alpha^{2}T^{00} - W\rho\\
        \alpha T^{0}_{\ i}
    \end{bmatrix}\;,
\label{eq:cons}
\end{equation}
where \mbox{$\gamma=\det\left(\gamma_{ij}\right)$} and \mbox{$W=\alpha u^{0}$} is the Lorentz factor. Note that $\rho_*$ is sometimes referred to in the literature as $\tilde{D}$. The adopted stress-energy tensor is that of a perfect fluid
\begin{equation}
    T^{\mu\nu} = \left(\rho h + b^{2}\right)u^{\mu}u^{\nu} + \left(P + P_\mathrm{mag}\right) g^{\mu\nu} - b^{\mu}b^{\nu}\;,
\end{equation}
where $h = 1 + \epsilon + P/\rho$ is the specific enthalpy, $\epsilon$ is the specific internal energy, \mbox{$g_{\mu\nu}$} is the four-metric, \mbox{$P_\mathrm{mag} = b^{2}/2$} is the magnetic pressure, \mbox{$b^{2} \equiv g_{\mu\nu}b^{\mu}b^{\nu}$} is the magnetic energy density, and $b^{\mu}$ is the magnetic field measured by an observer comoving with the fluid (i.e., with four-velocity $u^{\mu}$),
\begin{align}
    b^{0} &= u_{i}B^{i}/\alpha \;,\\
    b^{i} &= \left(B^{i}/\alpha + b^{0}u^{i}\right)/u^{0} = \left(B^i + B^j u_j u^i \right)/W\;.
\end{align}
% in the comoving frame of the fluid.

To complete our prescription of \eqref{eq:basic_evol}, the flux terms may be written as
\begin{equation}
    \fluxv^{j} =
    \begin{bmatrix}
        \rho_{*}v^{j}\\
        \yet v^{j}\\
        \tilde{S} v^{j}\\
        \alpha^{2}\sqrt{\gamma}T^{0j} - \rho_{*}v^{j}\\
        \alpha\sqrt{\gamma}T^{j}_{\ i}
    \end{bmatrix}\;,
    \label{eq:fluxes}
\end{equation}
and the source terms as
\begin{equation}
    \sourcev =
    \begin{bmatrix}
    0\\
    0\\
    0\\
    \alpha \sqrt{\gamma} \left[\Theta^{kl} K_{kl} - \left(T^{00}\beta^k + T^{0k} \right) \partial_k\alpha \right]\\
    \frac{1}{2} \alpha\sqrt{\gamma} T^{\mu\nu} g_{\mu\nu,i}
    \end{bmatrix}\;,
\end{equation}
where $\Theta^{kl} \equiv T^{00}\beta^k\beta^l + 2 T^{0k}\beta^l + T^{kl}$.

\section{Continuous Integration Testing}
\label{app:CI}

The industry standard for code testing is to have regularly scheduled tests which validate individual components, or units. These unit tests validate each individual component so that if a change breaks the code, it is clear which piece has diverged from the expected behavior. Since many codebases are worked on by large groups of people who may be implementing many different features simultaneously, it is standard to use a technique known as continuous-integration (CI) testing. CI runs unit tests every time the triggers are met, such as whenever someone pushes to the repository. The library is also indirectly tested via the \etk's CI, as several tests now include \grhayl-based thorns. However, we only discuss testing performed by the library and not third parties.

We perform CI testing via GitHub Actions. We trigger these tests every time a push is made to the \code{main} branch or when a pull request is opened or updated. A separate repository (\code{GRHayL\_TestData}) contains all the data needed to run the tests. The CI tests pull the individual data files they need using \code{curl}, run the \grhayl functions using the initial data, and then validate the output data against a trusted version from \origigm whenever possible. These tests are run on a variety of operating systems, versions, and compilers.

This method of automated unit testing is a required feature of most professional codes, and adopting these standards ensures that our code library continues to meet our expectations of accuracy and compatibility across systems, compilers, and infrastructures. As new code is introduced to \grhayl, the list of unit tests is expanded to validate the additions, ensuring that all the components of the library can be trusted to have identical behavior in the future (barring purposeful improvements or bug fixes which cause changes to behavior).

\subsection{Code Test Methodology}
\label{app:code_test_methodology}

\grhayl has been extensively tested, and automated tests immediately validate any changes to the code. The tests read input data $u_\mathrm{in}$ and compute an output $v_\mathrm{new}$. We compare this against a trusted reference output $v_\mathrm{old}$. To estimate round-off-level sensitivity, we additionally compute $v_\mathrm{pert}$ by perturbing $u_\mathrm{in}$ and rerunning the trusted code.

 We define the error bars as
\begin{equation}\label{eq:relerr}
    \left|\frac{v_\mathrm{new}-v_\mathrm{old}}{v_\mathrm{old}}\right| < 4\left|\frac{v_\mathrm{pert}-v_\mathrm{old}}{v_\mathrm{old}}\right|\;.
\end{equation}
The multiplication by 4 approximates the fact that the perturbed errors are unlikely to align to the maximum error. This error checking is very strict, far more than most codes demand for changes. Many codes set somewhat arbitrary error bars by using explicit relative and absolute tolerances that they hope correspond to the expected (worst-case) numerical error of the test. By estimating the numerical round-off error, we can more confidently state that the functions in \grhayl produce identical results to those produced by previous versions of the code.

As previously mentioned, the data used for these comparisons are kept in the \code{GRHayL\_TestData} repository. This serves to keep \grhayl as small as possible and also permits more and larger tests since we do not have to be as worried about the file sizes. These tests are optionally compiled along with the library, and the source code is in the \code{Unit\_Tests} directory. Any test that requires data from the \code{GRHayL\_TestData} repository also has a data generation file in the \code{data\_gen} subdirectory.

The tests can be separated into two categories. The tests prefaced with \code{ET\_Legacy} have matching code in \origigm. As such, the only data generated by \grhayl are the input data. A separate branch of \origigm (\code{GRHayLTestPatch}) contains the code changes and a parfile to read in this data. The branched \origigm code produces the trusted outputs $v_\mathrm{old}$ and $v_\mathrm{pert}$ as binary files for all the \code{ET\_Legacy} tests (which are also stored in the \code{GRHayL\_TestData} repository). The \grhayl library computes $v_\mathrm{new}$ from the same input data, and these values are compared to $v_\mathrm{old}$ and $v_\mathrm{pert}$, ensuring that the \grhayl code matches the progenitor code in \origigm. For new code, the data and data-generating code are grouped into directories by gem. The data-generating code for functions new to \grhayl produces input, output, and perturbed output binary files.

All tests (with one exception) perturb the input data with
\begin{equation}\label{eq:pertval}
u_\mathrm{pert} = u_\mathrm{in}\Bigl[1 + \mathrm{rand}\bigl(-10^{-14},10^{-14}\bigr)\Bigr]\;.
\end{equation}
By choosing such a small perturbation we demand a very high level of agreement across code versions, compilers, and operating systems. The benefit of using perturbed values as error bars is that we need not set a single tolerance cutoff for the test. Each data point in the test effectively has its own customized cutoff value. The common implementation of tolerance cutoffs requires a global tolerance value. This can, at best, match the error of the worst part of the variable space. Areas where the code should have greater accuracy can therefore be assessed, rather than sliding under the limit set by less accurate parts of the variable domain.

As mentioned, one test uses different perturbed input values. The comparison between the \origigm and \grhayl Con2Prim routines requires some relaxation of the stringent requirements on \grhayl tests. This is because the differences between the two are more significant than those found in other pieces of \origigm. Although \origigm is a long-standing and well-tested code, many aspects of the Con2Prim routine allow for extensive simplifications and reductions in computation. This results in many (accumulating) round-off-level differences before even reaching the Con2Prim routine itself. These effective perturbations are larger than the ``perturbed'' input data that we are using to approximate numerical error. These perturbations are further magnified by going through the Newton-Raphson method of Noble2D and several more variable conversions before returning to the top-level code. As such, we relax the error bars by changing the perturbation factor from $10^{-14}$ to $10^{-12}$ in \eqref{eq:pertval}. Another problem with this comparison is that the Noble2D Con2Prim routine used by both can have catastrophic cancellation under the right circumstances. This affects the computed pressure and can lead to apparent code failures. This only occurs for very small pressures, so this one test has, in addition to larger perturbed inputs (effectively a larger relative tolerance value), a larger absolute tolerance cutoff for the pressure.

\subsection{Code Coverage Tracking}
\label{app:code_coverage}

Another concern for high-quality code testing is ensuring that the code is truly being exercised. If a test is set up incorrectly but passes, it may not be obvious that the code is not, in fact, being tested. To remedy this, we use \code{gcovr} (the recursive version of \code{gcov}) to track which lines of code are covered by our tests. Since we have many tests covering the library, compiling a single report from all the individual \code{gcovr} reports would be quite tedious. We instead link our CI tests to Codecov, which collates all the data and provides a single report on code coverage, including the percentage of lines covered. This also includes coverage percentage information for subdirectories. At the level of individual files, \code{gcovr} even reports line-by-line code coverage information. This information is very helpful for designing future tests and making sure edge cases are being properly tested.

\bibliographystyle{apsrev4-1}
\bibliography{references}

\end{document}